\documentclass[final,5p,times,twocolumn]{elsarticle}

\usepackage{amssymb}
\usepackage{amsmath}

\usepackage{subfigure}
\usepackage{placeins} 
\usepackage{dblfloatfix}
\usepackage[colorlinks=true,linkcolor=blue,citecolor=blue,urlcolor=blue]{hyperref}  
\journal{}

\begin{document}

\begin{frontmatter}

\title{Structural Implications of the Chameleon Mechanism on White Dwarfs}

\author[CENTRA]{Joan Bachs-Esteban}
\author[CENTRA]{Ilídio Lopes}
\author[IPARCOS]{Javier Rubio}
 
\affiliation[CENTRA]{organization={Centro de Astrofísica e Gravitação - CENTRA, Departamento de Física, Instituto Superior Técnico - IST, Universidade de Lisboa - UL},
             addressline={Av. Rovisco Pais 1},
             postcode={1049-001},
             city={Lisboa},
             country={Portugal}}

\affiliation[IPARCOS]{organization={Departamento de Física Teórica and Instituto de Física de Partículas y del Cosmos (IPARCOS), Facultad de Ciencias Físicas, Universidad Complutense de Madrid},
             postcode={28040},
             city={Madrid},
             country={Spain}}

\cortext[cor]{Corresponding author: joan.bachs@tecnico.ulisboa.pt}
             
\begin{abstract}
We study the impact of the chameleon mechanism on the structure of white dwarfs.  Using a shooting method of our design, we solve the corresponding scalar-tensor equilibrium equations for a Chandrasekhar equation of state, exploring various energy scales and couplings of the chameleon field to matter. For the considered parameter ranges, we find the chameleon field to be in a thick-shell configuration, identifying for the first time in the literature a similarity relation of the theory for the radially normalised scalar field gradient. Our analysis reveals that the chameleon mechanism alters the internal pressure of white dwarfs, leading to a reduction in the stellar radii and masses and shifting the mass-radius curves below those predicted by Newtonian gravity. This lowers also the specific heat of white dwarfs, accelerating their cooling process. Finally, we derive parametric expressions from our results to expedite future analyses of white dwarfs in scalar-tensor theories.
\end{abstract}


\end{frontmatter}


\section{\label{sec:intro}Introduction}

General Relativity (GR) stands strong as a gravity theory, verified by numerous experiments and observations \cite{Will:2018bme}. Nevertheless, it fails to explain astrophysical and cosmological phenomena like the galactic rotation curves or the fundamental origin of dark energy. This suggests that gravity may not be fully described by GR but by a modified gravity (MG) theory \cite{Clifton:2011jh}.

Among the numerous extensions of GR, scalar-tensor (ST) theories of gravity \cite{Burrage_2018,Brax:2021wcv,Fischer:2024eic} 
stand out as one of the simplest yet elegant proposals. On general grounds, these types of settings introduce one or more scalar fields potentially mediating a fifth force among matter components, which, if sufficiently long-range, could potentially contradict the local tests of gravity \cite{Will:2018bme} or enhance structure formation in the early \cite{Amendola:2017xhl,Savastano:2019zpr,Goh:2023mau} and late Universe \cite{Wetterich:1994bg,Wetterich:2007kr,Amendola:2007yx,Casas:2016duf} (for notable exceptions see \cite{Garcia-Bellido:2011kqb,Ferreira:2016kxi,Casas:2018fum,Copeland:2021qby}). Many ST theories include, however, screening mechanisms that render the scalar field properties environment-dependent. In the chameleon mechanism \cite{Khoury_2004a}, the mass of the scalar field varies with the environment. In the symmetron \cite{Hinterbichler_2011} and dilaton scenarios \cite{Brax_2010}, it is the coupling to matter that changes, while in the Vainshtein \cite{Vainshtein_1972} and k-mouflage implementations \cite{Babichev_2009}, it is the kinetic function that plays a role. These screening mechanisms ensure that the scalar field becomes ineffective on astrophysical scales while remaining potentially relevant at cosmological scales.

The partial breaking of screening mechanisms within massive sources is expected to influence the equilibrium structure of stars, altering with it fundamental properties such as their mass-radius relations and cooling times (see~e.g.~\cite{Babichev:2009fi,Chang:2010xh,Sakstein_2013,Brito:2014ifa,Babichev:2016jom, de_Aguiar_2020, deAguiar_2021,Panotopoulos:2021rih,terHaar:2020xxb,Dima_2021}). Among the various compact objects suitable for studying the effects of screening mechanisms, white dwarfs (WDs) are particularly promising yet relatively unexplored targets \cite{Saltas_2018, Alam_2023, Kalita_2023, Vidal:2024wto}. This is due to two key reasons. Firstly, the equation of state (EoS) describing the microscopic behaviour of matter inside WDs is fairly well understood \cite{Camenzind_2007}. Secondly, the extensive observational data now available from \textit{Gaia}'s data releases \cite{Jim_nez_Esteban_2018, Tremblay_2018, Kilic_2020} provides a wealth of information on WDs, facilitating comprehensive studies of their spatial distribution, kinematics, and fundamental properties such as luminosity, temperature, and radius.

In this study, we focus on exploring the impact of the chameleon mechanism on the structure of WDs. To this end, we numerically solve the ST equilibrium equations in the relativistic and Newtonian limits and employ a Chandrasekhar EoS. By exploring a broad range of energy scales and conformal chameleon couplings to matter, we determine the corresponding mass-radius relations, deriving also a set of ready-to-use fitting formulae aiming to streamline future analyses of WDs in ST theories.

This paper is structured as follows: Section \ref{sec:theory} introduces the physical framework, discussing the ST theory and the equilibrium equations for static and spherically symmetric WDs together with the EoS for these stars. Section \ref{sec:Model} describes the specific screening mechanism under consideration and outlines the employed numerical methods, including boundary conditions and our customized shooting method. Section \ref{sec:Results} presents our findings regarding chameleon-screened WDs, followed by a discussion of implications and future directions in Section \ref{sec:Conclusion}. Finally, \ref{app:TOVvsNWT} provides supplementary details on the validity of the Newtonian approximation in our ST framework. 

In this work, we use the metric signature $(-,+,+,+)$ and consider $c=\hbar=k_B=1$ unless otherwise stated.

\section{\label{sec:theory}Framework}

\subsection{\label{subsec:ST} Scalar-Tensor Theory}

Several ST theories with environmentally dependent screening mechanisms such as chameleons, symmetrons, or dilatons can be described by a general action \cite{Sakstein_2013}
\begin{equation}\label{eq:STaction}
\begin{split}
    S=&\int d^4x\sqrt{-g}\left[\frac{M_P^2}{2}R - \frac{1}{2}\nabla_\mu\phi\nabla^\mu\phi - V(\phi)\right]\\
    &+ S_m\left[\Psi_m;A^2(\phi)g_{\mu\nu}\right],
\end{split}
\end{equation}
with $M_P=(8\pi G)^{-1/2}=2.43\times 10^{18}$ GeV the reduced Planck mass, $g$ and $R$ the determinant and Ricci scalar of the \textit{Einstein frame} metric $g_{\mu\nu}$ and $\phi$ a scalar field. Each model belonging to this class of theories is characterised by a self-interacting potential $V(\phi)$ and a conformal coupling $A(\phi)$ to the matter fields $\Psi_m$. In particular, the scalar field is taken to be gravitationally coupled to the matter fields through a conformally rescaled \textit{Jordan frame} metric $\Tilde{g}_{\mu\nu}\equiv A^2(\phi)g_{\mu\nu}$. This modifies the Newtonian force in the non-relativistic limit, making these ST theories MG theories \cite{Sakstein_2013}. 
 
By varying Eq.~\eqref{eq:STaction} with respect to the metric, one obtains the field equations
\begin{equation}\label{eq:EF_eq}
    G_{\mu\nu}=\kappa^2\left[T_{\mu\nu}+\nabla_\mu\phi\nabla_\nu\phi-g_{\mu\nu}\left(\frac{1}{2}\nabla_\sigma\phi\nabla^\sigma\phi+V(\phi)\right)\right]\,,
\end{equation}
where $G_{\mu\nu}$ is the Einstein tensor, $\kappa\equiv M_P^{-1}$, and \begin{equation}\label{eq:EF_Tensor_Def}
    T_{\mu\nu}\equiv-\frac{2}{\sqrt{-g}}\frac{\delta S_m}{\delta g^{\mu\nu}}
\end{equation}
is the energy-momentum tensor of the matter fields, which we assume to be described by a perfect fluid, that is
\begin{equation}\label{eq:EF_Tensor_Perfect_Fluid}
    T^{\mu\nu}\equiv(\epsilon+P)u^\mu u^\nu+Pg^{\mu\nu}\,,
\end{equation}
with $u^\mu$ the four-velocity of fluid elements, and $\epsilon$ and $P$ the total energy density and pressure in the fluid's rest frame, respectively. Analogously, if we vary Eq.~\eqref{eq:STaction} with respect to the field, we obtain the scalar field equation
\begin{equation}\label{eq:phi_eq}
    \Box\phi=\frac{dV(\phi)}{d\phi}-\frac{d\text{ ln }A(\phi)}{d\phi}T\equiv\frac{dV_{\rm eff}(\phi)}{d\phi}\,,
\end{equation}
with $V_{\rm eff}(\phi)$ the potential effectively governing $\phi$.
The matter equation of motion is determined by the divergence of Eq.~\eqref{eq:EF_eq}, namely
\begin{equation}\label{eq:EF_Matter_EOM}
    \nabla^\nu T_{\mu\nu}=\frac{d\text{ ln }A(\phi)}{d\phi}T\nabla_\mu\phi\,,
\end{equation}
with $T\equiv g^{\mu\nu}T_{\mu\nu}$ denoting the trace of the energy-momentum tensor. The above equation means that particles do not follow geodesics in the Einstein frame metric $g_{\mu\nu}$, being their trajectories also affected by the scalar field gradient. Nevertheless, in an alternative, but equally valid description, we can analogously define the Jordan frame matter energy-momentum tensor as
\begin{equation}\label{eq:JF_Tensor_Def}
    \Tilde{T}_{\mu\nu}\equiv-\frac{2}{\sqrt{-\Tilde{g}}}\frac{\delta S_m}{\delta \Tilde{g}^{\mu\nu}}\,.
\end{equation}
Comparing the latter expression with Eq.~\eqref{eq:EF_Tensor_Def}, we see that both tensors are related through $T_{\mu\nu}=A^2(\phi)\Tilde{T}_{\mu\nu}$. Using Eq.~\eqref{eq:STaction}, one can show that $\Tilde{T}_{\mu\nu}$ is indeed covariantly conserved -- i.e., $\Tilde{\nabla}^\nu\Tilde{T}_{\mu\nu}=0$ -- and that free particles follow the $\Tilde{g}_{\mu\nu}$ geodesics. In addition, from the four-velocity normalisation condition $g_{\mu\nu} u^\mu u^\nu =-1$, we get the relation $u^\mu=A(\phi)\Tilde{u}^\mu$. This conformal transformation, together with that for the energy-momentum tensor above, allows us to find the correspondence between the fluid variables in both frames, namely $\epsilon=A^4(\phi)\Tilde{\epsilon}$ and $P=A^4(\phi)\Tilde{P}$.

\subsection{\label{subsec:EoS}Equation of State}

The equation of state (EoS) condenses the microphysics of the stellar interior in a relation between pressure and density. As explained now and discussed in \ref{app:TOVvsNWT}, WDs can be adequately described as non-relativistic objects, both in GR and the ST scenarios considered here. Therefore, it is useful to introduce the rest-mass density $\Tilde{\rho}$ and the internal energy density $\Tilde{\Pi}$, which are related to the total energy density $\Tilde{\epsilon}$ as
\begin{equation}\label{eq:rest_mass_density}
    \Tilde{\epsilon}\equiv \Tilde{\rho}\left(1+\frac{\Tilde{\Pi}}{\Tilde{\rho} c^2}\right)\,,
\end{equation}
where we have explicitly written the speed of light $c$ to evince that $\Tilde{\Pi}$ is a first-order relativistic correction.

Moreover, the pressure and energy in WDs are nonthermal, in the sense that thermal effects can be modelled as small perturbations on top of the fluid dynamics \cite{Camenzind_2007}. Consequently, the EoS reduces to single parameter functions, namely $\Tilde{P}(\Tilde{\rho})$ and $\Tilde{\epsilon}(\Tilde{\rho})$. Notice that we are relating the EoS to the Jordan-frame variables, as the common thermodynamic relation for energy conservation, $d(\Tilde{\epsilon}/\Tilde{\rho})=-\Tilde{P}d(1/\Tilde{\rho})$, holds only in this frame. 

In WDs, the electrostatic energy of the matter structure is negligible as compared to the Fermi energies, hence the Coulomb forces are too. Therefore, the electron pressure is given by \cite{Camenzind_2007}
\begin{equation}\label{eq:Electron_pressure}
    \Tilde{P}=\frac{2}{(2\pi)^3 \hbar^3}\int_0^{p_{F,e}}\frac{p^2c^2}{\sqrt{p^2c^2+\left(m_ec^2\right)^2}}4\pi p^2dp=\frac{m_ec^2}{\lambda_e^3}\psi(x)\,,
\end{equation}
with
\begin{equation}\label{eq:Electron_pressure_formula}
    \psi(x)=\frac{1}{8\pi^2}\left[x\sqrt{1+x^2}\left(\frac{2x^2}{3}-1\right)+\text{ln}\left(x+\sqrt{1+x^2}\right)\right],
\end{equation}
where, for the sake of clarity, we have made explicit again the different $\hbar$ and $c$ factors. The factor 2 in Eq.~\eqref{eq:Electron_pressure} is due to the electron spin degeneracy, $p_F$ is the Fermi momentum of the electrons, $m_e$ is the electron mass, $\lambda_e\equiv \hbar/(m_e c)$ is the electron Compton wavelength, and $x\equiv p_F/m_ec$ is the dimensionless Fermi momentum.

Even though the main contribution to WD pressure comes from the degenerate electrons, the energy is dominated by the ions. Since these are non-relativistic for densities below the neutron drip $\Tilde{\rho}_{\text{n-drip}}\simeq4\times10^{11}\text{ g}\,\text{cm}^{-3}$ \cite{Camenzind_2007}, the energy density can be expressed in terms of the rest-mass density
\begin{equation}
    \Tilde{\epsilon}=\Tilde{\rho}=\frac{m_Bn_e}{Y_e}\,,
\end{equation}
where $m_B$ is the mean nucleon mass, $n_e=8\pi p_{F,e}^3/(3h^3)$ is the electron number density, and $Y_e = Z/A$ (with $Z$ the atomic number and $A$ the atomic weight) is the mean number of electrons per nucleon. WDs are usually modelled as a cold, degenerate matter star made of helium, carbon, or oxygen \footnote{The first EoS for such stars was derived by Chandrasekhar \cite{Chandrasekhar:1935zz}. Hamada and Salpeter added temperature corrections to it \cite{1961ApJ...134..683H}.}. For any of these elements, $Y_e =0.5$ when they are fully ionised \cite{Camenzind_2007}. The mean nucleon mass of carbon is $m_{B,C}=1.66057\times10^{-24}\,\text{g}$. Thus, we will express the density $\Tilde{\rho}$ and the pressure $\Tilde{P}$ as
\begin{eqnarray}
    \Tilde{\rho}(x)=&&1.9479\times10^6\,x^3\,\text{g}\,\text{cm}^{-3}\,,\label{eq:EoS_rho}\\
        \Tilde{P}(x)=&&1.4218\times10^{25}\,\psi(x)\,\text{dyn}\,\text{cm}^{-2\label{eq:EoS_P}}\,.
\end{eqnarray}

Although more precise EoS exist to describe the internal structure of a WD \cite{Camenzind_2007}, considering the exploratory nature of this work and the advantage of simplicity, we opted to use this equation for a more efficient calculation method. Moreover, using a more precise EoS would not significantly alter our results and conclusions.

\subsection{\label{subsec:EqEq}Equilibrium Equations}

In the Newtonian description, the gravitational field is weak and static and particles move slowly compared to the speed of light \cite{Misner_1973}, which means that pressure is negligible with respect to energy density. Thus, the WD line element can be written as
\begin{equation}\label{eq:metric_spherical_polar_NewtonLimit}
    ds^2=-\left(1+2\Phi(r)\right)dt^2+\left(1-2\Phi(r)\right)dr^2+r^2d\Omega^2\,,
\end{equation}
with $\Phi(r)$ the Newtonian potential and $d\Omega^2=d\theta^2+\text{sin}^2\theta d\varphi^2$ is the two-sphere line element. Replacing this metric into Eqs.~\eqref{eq:EF_eq}, \eqref{eq:EF_Matter_EOM}, and \eqref{eq:phi_eq}, one has that
\begin{eqnarray}
    \Phi'=&&\frac{m}{r^2}\,,\label{eq:background_Phi_Newton}\\
    m'=&&\frac{\kappa^2}{2}r^2A^4\Tilde{\rho}\,,\label{eq:background_m_Newton}\\
    \Tilde{P}'=&&-\Tilde{\rho}\left(\Phi'+\frac{A_{,\phi}}{A}\sigma \right)\,,\label{eq:background_p_Newton}\\
    \phi'=&&\sigma\,,\label{eq:background_phi_Newton}\\
    \sigma'=&&-\frac{2}{r}\sigma+V_{,\phi}+A_{,\phi}A^3\Tilde{\rho}\,,\label{eq:background_sigma_Newton}
\end{eqnarray}
where we have ignored pressure contributions as to energy contributions and neglected second-order terms. We also have replaced $\Tilde{\epsilon}$ with $\Tilde{\rho}$ since these two quantities are equivalent in the Newtonian limit (recall Eq.~\eqref{eq:rest_mass_density}).

Once one chooses the model functions $V(\phi)$ and $A(\phi)$, and a suitable EoS, the system of differential of ordinary differential equations (ODE) can be numerically integrated from the origin. For each central density $\Tilde{\rho}_0=\Tilde{\rho}(0)$, we obtain the stellar mass $M$ and the stellar radius $R$. Therefore, if we perform the integration for a range of central densities, we can get a so-called mass-radius (MR) curve. This family of stars is unique for each EoS and is parametrised by the central density $\Tilde{\rho}_0$ \cite{Camenzind_2007}. We provide further details of the integration process and the boundary conditions in Sec.~\ref{subsec:BC}. 

\section{\label{sec:Model}Model}

\subsection{\label{subsec:Screening}Chameleon Screening}

In this work, we focus on the chameleon field \cite{Khoury_2004a}, a scalar field equipped with a screening mechanism through its effective mass. The environment-dependent mass of the chameleon comes from the synergy between the self-interacting potential $V(\phi)$ and the conformal coupling $A(\phi)$. The potential should be monotonically decreasing and of runaway form, such that it does not have a minimum but that, together with the coupling, they bestow a minimum to the effective potential $V_{\rm eff}(\phi)$. We then consider a classic inverse power-law potential and an exponential conformal coupling
\begin{equation}\label{eq:potential_&_conformal_coupling}
    V(\phi)=\Lambda^4\left(\frac{\Lambda}{\phi}\right)^n\,,\quad \quad A(\phi)=e^{\beta\phi/M_P}\,,
\end{equation}
where $n$ is a positive constant, $\Lambda$ has mass units, and $\beta$ is a dimensionless constant. The $\beta$ parameter is the \textit{coupling strength} between the scalar and matter fields, and $\Lambda$ controls the scalar field contribution to the energy density of the universe, hence we shall refer to it as the \textit{chameleon energy scale}. For $n$ and $\beta$ of order unity, equivalence principle tests impose that $\Lambda\lesssim10^{-30}\,M_P\approx 1\,\text{meV}$ \cite{Khoury_2004a}, which remarkably coincides with the dark energy scale causing the current accelerated expansion of the universe. Nevertheless, we do not regard the chameleon field studied in this work as the force driving the cosmological expansion. 

The minima $\Bar{\phi}$ of the chameleon field are the roots of Eq.~\eqref{eq:phi_eq}, which are determined by the transcendental equation $nM_P\Lambda^{n+4} + \beta\Tilde{T}\Bar{\phi}^{n+1}e^{4\beta\Bar{\phi}/M_P}=0$ since the traces of the energy-momentum tensor in the Einstein and Jordan frames are related through the expression $T=A^4(\phi)\Tilde{T}$. In the $\beta\Bar{\phi}/M_P\ll 1$ limit, we can approximate the solution  by
\begin{equation}\label{eq:Veff_minimum_value}
    \Bar{\phi}\approx\sqrt[n+1]{\frac{nM_P\Lambda^{n+4}}{-\beta\Tilde{T}}}\approx\sqrt[n+1]{\frac{nM_P\Lambda^{n+4}}{\beta\Tilde{\rho}}}\,,
\end{equation}
which will be real whenever $\Tilde{T}<0$ since $n$, $\Lambda$, and $\beta$ are positive. This is the case in the Newtonian limit, for the trace is $\Tilde{T}=3\Tilde{P}-\Tilde{\rho}$ and pressure is negligible in front of energy density, thus one has that $\Tilde{T}\approx-\Tilde{\rho}$, hence the second expression.

Let us study how the chameleon fifth force is screened in WDs. For explanatory purposes, we consider a static, spherically symmetric WD of total mass $M$, density $\Tilde{\rho}_0$, and radius $R$, surrounded by a medium whose density $\Tilde{\rho}_\infty$ is much smaller than that of the star -- i.e.,  $\Tilde{\rho}_\infty\ll\Tilde{\rho}_0$ -- for instance, the cosmological background. From Eq.~\eqref{eq:Veff_minimum_value}, we deduce that the chameleon will set to a minimum value within the star, $\Bar{\phi}_0$, that will be lower than the minimum outside of it, $\Bar{\phi}_\infty$, since $\Bar{\phi}$ is inversely proportional to the environment density. 

One calculates the chameleon's mass of small fluctuations around a potential minimum $\Bar{\phi}$ by evaluating the second derivative of the effective potential with respect to the scalar field at $\Bar{\phi}$. Therefore, deriving Eq.~\eqref{eq:phi_eq} and considering the chameleon model in \eqref{eq:potential_&_conformal_coupling}, we have that
\begin{equation}\label{eq:EffectiveMass}
    m_{{\rm eff}}^2\equiv\frac{d^2V_{{\rm eff}}}{d\phi^2}=n(n+1)\frac{\Lambda^{n+4}}{\phi^{n+2}}-4\frac{\beta^2}{M_P^2}e^{4\beta\phi/M_P}\Tilde{T}\,.
\end{equation}
Then, when we replace the scalar field value for the expression in Eq.~\eqref{eq:Veff_minimum_value}, we see that the chameleon's effective mass for a WD is proportional to the density since it is a non-relativistic object
\begin{equation}\label{eq:EffectiveMassMinimum}
    m_{{\rm eff}}^2\vert_{\Bar{\phi}}\approx\frac{n+1}{\sqrt[n+1]{n\Lambda^{n+4}}}\left(\frac{\beta}{M_P}\Tilde{\rho}\right)^{\frac{n+2}{n+1}}\,.
\end{equation}
We have considered the approximation $\beta\Bar{\phi}/M_P\ll 1$ and neglected the $\beta^2$ term since the typical densities for WDs are much smaller than the Planck scale, so $\beta^2\Tilde{\rho} / M_P^2 \ll 1$. We can estimate the appropriate chameleon energy scale for a screened WD using Eq.~\eqref{eq:EffectiveMassMinimum} and imposing that the interaction range of the scalar field is of the size of the star. For a typical WD of radius $R\sim10^4$ km, central density $\Tilde{\rho}_0 \sim 10^6 \text{ g}\,\text{cm}^{-3}$, and $n=\beta=1$, one has that $\Lambda\sim10^{-18}\,M_P$. We explore intervals around these reference values in the numerical results of Sec.~\ref{sec:Results}.

Since the interaction range of the scalar field is inversely proportional to the effective mass, Eq.~\eqref{eq:EffectiveMassMinimum} means that the chameleon fifth force will be short-range in dense environments -- like the one under consideration -- and will be acting as a long-range force on cosmological scales. For instance, the effective mass of the chameleon inside the star will be much higher than the chameleon mass at cosmological scales, that is $m_{{\rm eff},0}\gg m_{{\rm eff},\infty}$. This is the key to the screening mechanism.

Qualitatively, one can distinguish two different screening regimes according to the behaviour of the field inside the star \cite{Khoury_2004a}. In the so-called \textit{thin-shell} regime, the chameleon field remains approximately constant within the star, changing only in a very thin region close to the stellar radius. On the contrary, in the alternative \textit{thick-shell} regime, the scalar field evolves right from the very centre of the star. In the latter situation, the solution for the scalar field can be approximately written as
\begin{align}
    \phi(r)\approx&\frac{\beta\rho_cr^2}{6M_P}+\Bar{\phi}_0,\hspace{2.85 cm} 0<r<R,\label{eq:thick_inside}\\
    \phi(r)\approx&-\frac{\beta}{4\pi M_P}\frac{Me^{-m_{{\rm eff},\infty}(r-R)}}{r}+\Bar{\phi}_\infty,\quad r>R.\label{eq:thick_outside}
\end{align} 
Note that, while capturing the essence of the thick-shell regime, these analytical expressions are based on three assumptions, which are not guaranteed to be satisfied in the problem under consideration.
First, the star of mass $M$ and radius $R$ is taken to have a homogeneous density $\rho_c$. Second, the contribution of the potential $V$ is assumed to be negligible as compared to that of the coupling $A$ inside the star. Last, the scalar field gradient is required to be large enough as compared to the curvature of the potential outside the star.

\subsection{\label{subsec:BC}Boundary Conditions}

Note that we do not need to specify boundary conditions for the gravitational potential $\Phi(r)$ -- as long as we are in an equilibrium configuration -- since the system of equations \eqref{eq:background_Phi_Newton}-\eqref{eq:background_sigma_Newton} depends only on its radial derivatives. For the pressure, we set $\Tilde{P}(0)=\Tilde{P}_0$, where $\Tilde{P}_0$ is the pressure at the WD centre. In our numerical integration, this value will cover a range of pressures belonging to the EoS validity domain. Actually, we will consider a range of central densities and calculate the corresponding central pressures through the EoS introduced in Sec.~\ref{subsec:EoS} (Eqs.~\eqref{eq:EoS_rho} and \eqref{eq:EoS_P}). The central densities we will employ are $\Tilde{\rho}_0 = 7\times10^{4} - 10^{10} \text{ g}\,\text{cm}^{-3}$, which render WDs with masses $M = 0.12 - 1.42 \,M_\odot$ and radii $R=1.3 - 17.3\text{ km}$ in Newtonian gravity, typical values for this kind of stars. Regarding the mass, one should set $m(0)=0$. However, since our code starts from a certain initial radius $r=r_0>0$, at which we consider the density to be $\Tilde{\rho}_0$, the initial condition of the mass is $m(r_0)=(4/3)\pi r_0^3\Tilde{\rho}_0$.    

We assume that WDs lie within a galaxy of density $\Tilde{\rho}_G=10^{-24} \text{ g}\,\text{cm}^{-3}$ \cite{Khoury_2004a}, meaning $\Tilde{\rho}=\Tilde{\rho}_G$ outside the star. This condition is not only necessary because we have required spacetime to become Schwarzschild far from the star, but also because we need a background density outside the star for the chameleon to achieve an effective potential minimum at infinity. Therefore, the stellar radius $R$ is determined by the condition $\Tilde{\rho}(R)=\Tilde{\rho}_\infty=\Tilde{\rho}_G$, although numerically this will translate into $\Tilde{\rho}(R)$ being close to $0$ given a certain tolerance since $\Tilde{\rho}_G$ is very small. In practice, we integrate up to a certain distance, which is big enough compared to the typical WDs radii that it can be thought of as infinite, and then we look for the radial coordinate at which the density is smaller than the tolerance. That coordinate is the stellar radius $R$, and the stellar mass $M$ is defined as the total mass within $R$ following Eq.~\eqref{eq:background_m_Newton}. 

Since the solution for the scalar field must be regular at the centre of the star, we know that the scalar field gradient fulfils $\sigma(0)=0$. We cannot know the value of the scalar field at the centre before the integration, but we do know that at infinity it will reach the exterior minimum, that is $\phi\rightarrow\Bar{\phi}_\infty$ as $r\rightarrow\infty$. This condition implies that the solution is also regular at infinity, i.e. $\sigma\rightarrow0$ as $r\rightarrow\infty$. Thus, to solve this ODE system, we ought to implement a shooting method, so we can find out the adequate value of $\phi$ at $r=0$ that leads to $\Bar{\phi}_\infty$ when we are away from the star. 

\subsection{\label{subsec:SM}Shooting Method}

The scalar field $\phi$ is governed by a second-order differential equation, Eq.~\eqref{eq:phi_eq}, which we have split into two first-order differential equations, Eqs.~\eqref{eq:background_phi_Newton} and \eqref{eq:background_sigma_Newton}. As explained in Sec.~\ref{subsec:BC}, we know both boundary conditions for the scalar field gradient $\sigma$, but we only know the infinity one for $\phi$, leaving the one at the origin to find.

We can estimate the value of the scalar field at the centre of the WD -- let us call it $\phi_0$ -- from Eq.~\eqref{eq:Veff_minimum_value}. This is possible because WDs are Newtonian astrophysical objects and, as we explained in Sec.~\ref{subsec:Screening}, the chameleon effective potential has a minimum inside such a star. Thus, we take it as a sensible guess. We then perform the numerical integration of the ODE system, either the relativistic (Eqs.~\eqref{eq:background_nu}-\eqref{eq:background_sigma}) or the Newtonian one (Eqs.~\eqref{eq:background_Phi_Newton}-\eqref{eq:background_sigma_Newton}). Afterwards, we compute the relative error between the scalar field minimum at infinity, $\Bar{\phi}_\infty$, and the value provided by our code, $\phi(r_{\text{max}})$, where $r_{\text{max}}$ is the maximum radial coordinate. If the relative error is smaller than a given tolerance -- that is, if $|\Bar{\phi}_\infty - \phi(r_{\text{max}})| / \Bar{\phi}_\infty < \phi_{\text{tol}}$ -- we have achieved convergence and, consequently, we store the output. 

If the tolerance criterion is not met, we increase or decrease $\phi_0$ by a small amount $\delta\phi$ depending on whether the difference between the theoretical and the computed value -- that is $\Bar{\phi}_\infty - \phi(r_{\text{max}})$ -- is positive or negative. At every step of the shooting method, we check the sign of the mentioned difference and, whenever it changes (indicating that we have gone beyond the desired value), we reduce $\delta\phi$. In this way, we boost convergence and achieve higher precision in fewer steps.

\section{\label{sec:Results}Results}

In this section, we present the results we have obtained from the numerical integration of Eqs.~\eqref{eq:background_Phi_Newton}-\eqref{eq:background_sigma_Newton} with the model functions $V(\phi)$ and $A(\phi)$ from \eqref{eq:potential_&_conformal_coupling}, considering the EoS given by Eqs.~\eqref{eq:EoS_rho} and \eqref{eq:EoS_P}, and using the shooting method detailed in Sec.~\ref{subsec:SM}. We have considered central densities ranging from $\Tilde{\rho}_{0,\text{min}} = 7\times10^{4} \text{ g}\,\text{cm}^{-3}$ to $\Tilde{\rho}_{0,\text{max}} = 10^{10} \text{ g}\,\text{cm}^{-3}$ and a background density $\Tilde{\rho}_\infty=10^{-4}\Tilde{\rho}_{0,\text{min}}$. For each $\Tilde{\rho}_0$ value, we obtain the radius $R$ and the mass $M$ of the WD, as we discussed in Sec.~\ref{subsec:EqEq}.

We consider $n=1,2$, coupling strengths $\beta = 0.1,\, 0.05,\, 0.01$, and energy scales between $\Lambda = 10^{-19}\,M_P$ and $\Lambda = 1.5\times10^{-18}\,M_P$. Let it be noted that $\Lambda\lesssim10^{-30}\,M_P$ should be reached to satisfy equivalence principle constraints \cite{Khoury_2004a}, but reaching such small values is numerically very expensive. In our shooting method, we set $\phi_{\text{tol}}=10^{-10}$, and to achieve such precision we already need to work with a considerable amount of significant digits, even for $\Lambda \sim 10^{-18}\,M_P$. Nonetheless, our main conclusions would also stand for realistic values of $\Lambda$, as we shall discuss.

\subsection{\label{subsec:Stellar}Stellar Structure}

\begin{figure}[hbt!]
\includegraphics[width=\columnwidth]{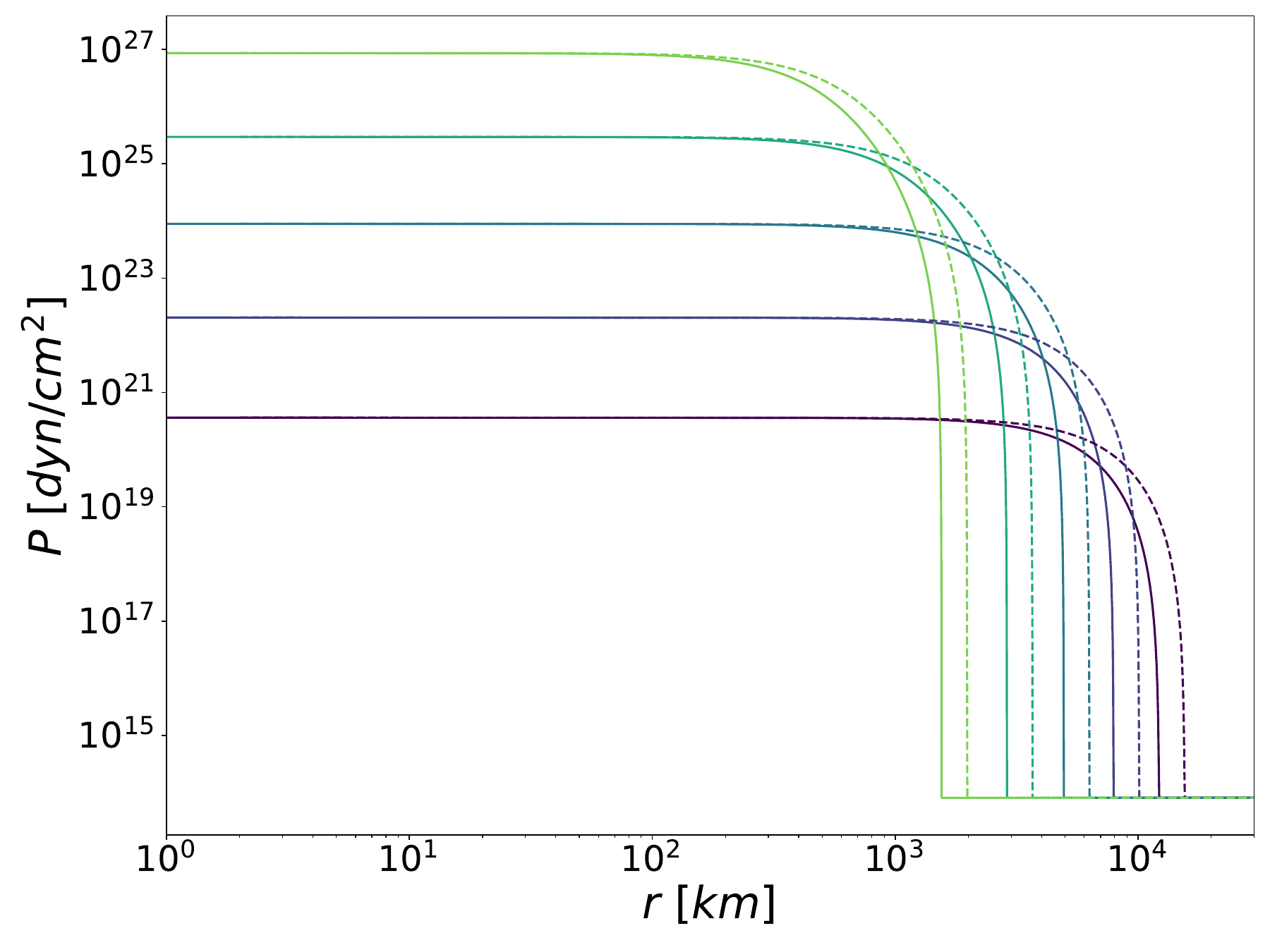}
\caption{Pressure radial profiles of chameleon-screened WDs. We consider the chameleon model \eqref{eq:potential_&_conformal_coupling} with $n=1$, $\Lambda=10^{-18}\,M_P$, and $\beta = 0.1$ (solid) and $\beta = 0.01$ (dashed). Colours from purple to green indicate increasing central densities, specifically for $\Tilde{\rho}_0 = 7.0\times10^{4},\,8.5\times10^{5},\,1.0\times10^{7},\,1.2\times10^{8},\,1.5\times10^{9}\text{ g}\,\text{cm}^{-3}$. The lower the coupling strength $\beta$, the longer it takes for the pressure to decrease.}
    \label{fig:Pvsr_L1e18}
\end{figure}

Fig.~\ref{fig:Pvsr_L1e18} shows the pressure radial profiles $\Tilde{P}(r)$ for WDs in two different realisations of the chameleon model characterised by different coupling strengths, namely $\beta = 0.1$ and $\beta = 0.01$, while fixing the other two parameters to $n=1$ and $\Lambda=10^{-18}\,M_P$. As for a star in GR or Newtonian gravity, the pressure $\Tilde{P}$ decreases along the radius. More precisely, it drops when $r\sim10^3-10^4$ km, which is the range of WDs radii. We observe that the lower the coupling strength $\beta$, the longer it takes for the pressure to decrease. From a mathematical point of view, this is easily understood from Eq.~\eqref{eq:background_p_Newton}. For our choice of chameleon functions, the term $A_{,\phi}/A$ is simply $\beta/M_P$. Hence, the rate at which the pressure diminishes is directly proportional to the product of $\beta$ and $\sigma$. 

if the coupling between the scalar field and matter is weaker, with positive coupling defined here.

From a physical perspective, it is obvious that the pressure decrease will be less affected by the scalar field if the coupling between the latter and the matter is weaker, with positive coupling defined here. Regarding the scalar field gradient $\sigma$, we know it will also be positive. The scalar field has a minimum inside the star and another one outside of it, the latter being higher than the former since the outside density is lower than the inside one (recall Eq.~\eqref{eq:Veff_minimum_value}). Plus, since there are no other extrema between these two minima, the scalar field always increases. So,  $\beta\sigma$ will always be positive (see Figs.~\ref{fig:phi&sigmavsrR_L5e19_a5e2} and \ref{fig:phi&sigmavsrR_L5e19_a1e2} for computational evidence). Then, since the hydrostatic equation \eqref{eq:background_p_Newton} has a global negative sign, the scalar contribution to it will always boost the pressure decrease.

As we can already imagine, this pressure drop will cause the chameleon-screened WDs to have different masses and radii than the ones in GR or Newtonian gravity. Since the pressure blue-- and therefore the density -- falls earlier, we achieve the condition $\Tilde{\rho}(R)=\Tilde{\rho}_\infty$ sooner, thus we get a smaller stellar radius $R$ in chameleon screening than in GR. This reduction is translated also to the stellar mass since it is defined as the mass contained in $R$. Accordingly, we obtain less massive stars in our chameleon model (see Sec.~\ref{subsec:MassRadius}).

\subsection{\label{subsec:CoolTime}Cooling Time}

The presence of the chameleon field is expected to affect the thermal properties of WDs. To quantify this, we consider the mean specific heat 
\begin{equation}\label{eq:specific_heat}
    \Bar{c}_V=\frac{1}{M}\int_0^M\left(c_V^{\text{ion}}+c_V^{\text{el}}\right)dm\,,
\end{equation}
where $M$ is the stellar mass, $c_V^{\text{ions}}$ is the specific heat of ions, and $c_V^{\text{el}}$ that for the electrons \cite{Kalita_2023}. The former depends on the ratio of Coulomb to thermal energy $\Gamma$, whose critical value is around $\Gamma_c=125$ \cite{Brush_1966}. If $\Gamma < \Gamma_c$, the specific heat of ions is constant, namely $c_V^{\text{ion}}=(3/2)k_B=3/2$, and if $\Gamma > \Gamma_c$, it depends on the temperature as 
\begin{equation}\label{eq:ion_heat}
    c_V^{\text{ion}}=9\left(\frac{T}{\Theta_D}\right)^3\int_0^{\Theta_D/T}\frac{x^4e^x}{(e^x-1)^2}dx\,,
\end{equation}
where 
\begin{equation}\label{eq:Debye_T}
    \Theta_D=3.48\times10^3 Y_e\sqrt{\Tilde{\rho}}
\end{equation}
stands for the Debye temperature in $K$, with the stellar density expressed in $\text{g}\,\text{cm}^{-3}$. 
The electrons' specific heat depends also on temperature through the expression \cite{Koester_1972}
\begin{equation}\label{eq:e_heat}
    c_V^{\text{el}}=\frac{\pi^2}{2}Z\frac{T}{\Tilde{\epsilon}_F}\,,
\end{equation}
where $\Tilde{\epsilon}_F = \Tilde{p}_F^2 + m_e^2$ is the Fermi energy and $p_F^3=3\pi^2Y_e\Tilde{\rho}/m_p$ is the Fermi momentum (recall Sec.~\ref{subsec:EoS}), with $m_p$ the proton mass.

\begin{figure}[hbt!]
\includegraphics[width=\columnwidth]{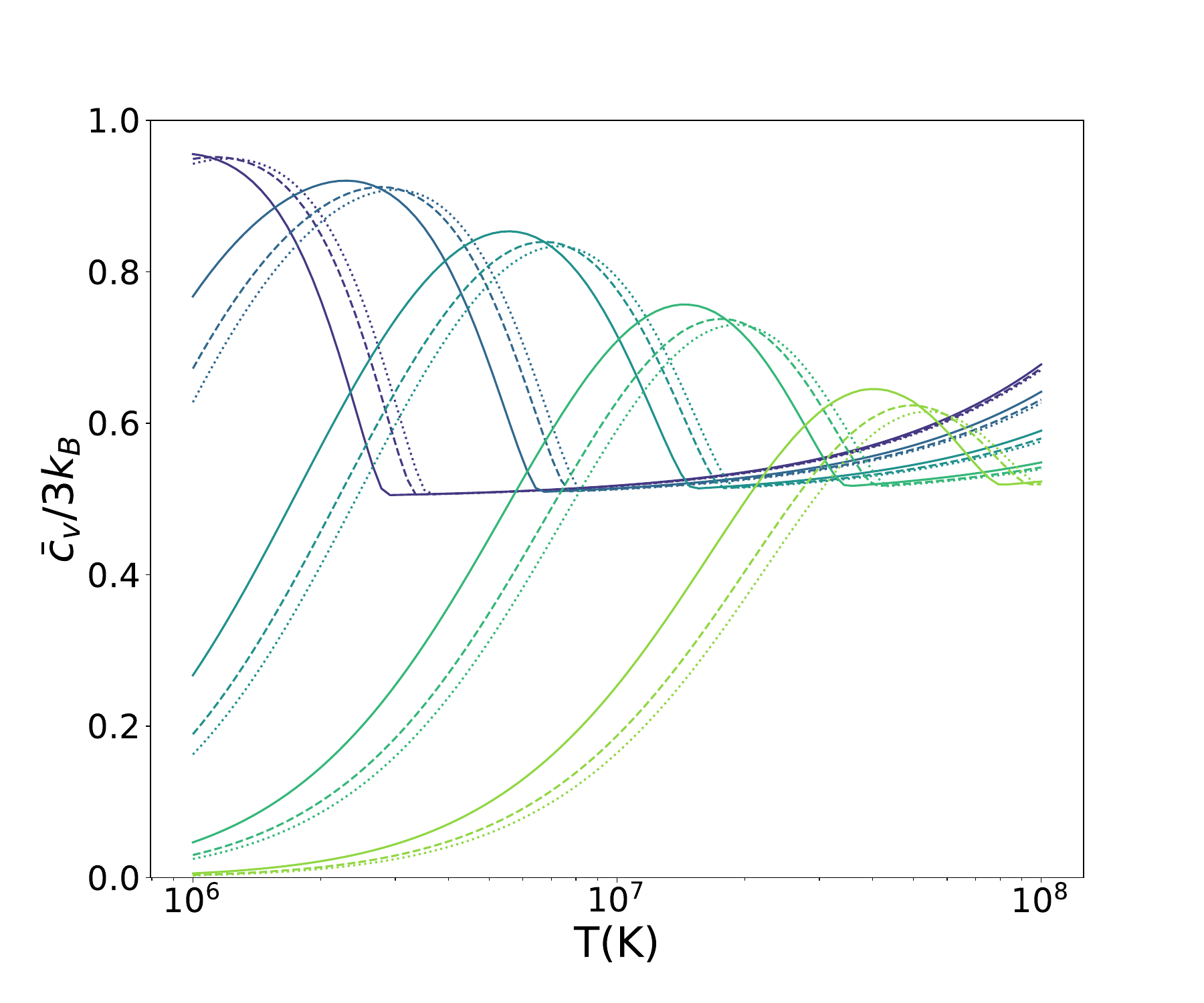}
\caption{Mean specific heat $\Bar{c}_V$ as a function of temperature $T$ for chameleon-screened WDs. We consider the chameleon model \eqref{eq:potential_&_conformal_coupling} with $n=1$, $\Lambda=10^{-18}\,M_P$, and $\beta = 0.1$ (dotted) and $\beta = 0.05$ (dashed). For reference, we include the results for WDs in Newtonian gravity (solid). Colours from purple to green indicate increasing central densities, specifically $\Tilde{\rho}_0 = 4.5\times10^{5},\,5.5\times10^{6},\,6.7\times10^{7},\,8.2\times10^{8},\,1.0\times10^{10}\text{ g}\,\text{cm}^{-3}$. For higher values of the coupling strength $\beta$, the $\Bar{c}_V$ maximum decreases and the entire curve shifts to higher temperatures.} \label{fig:c_VvsT_L1e18}
\end{figure}

\begin{figure}[hbt!]
\includegraphics[width=\columnwidth]{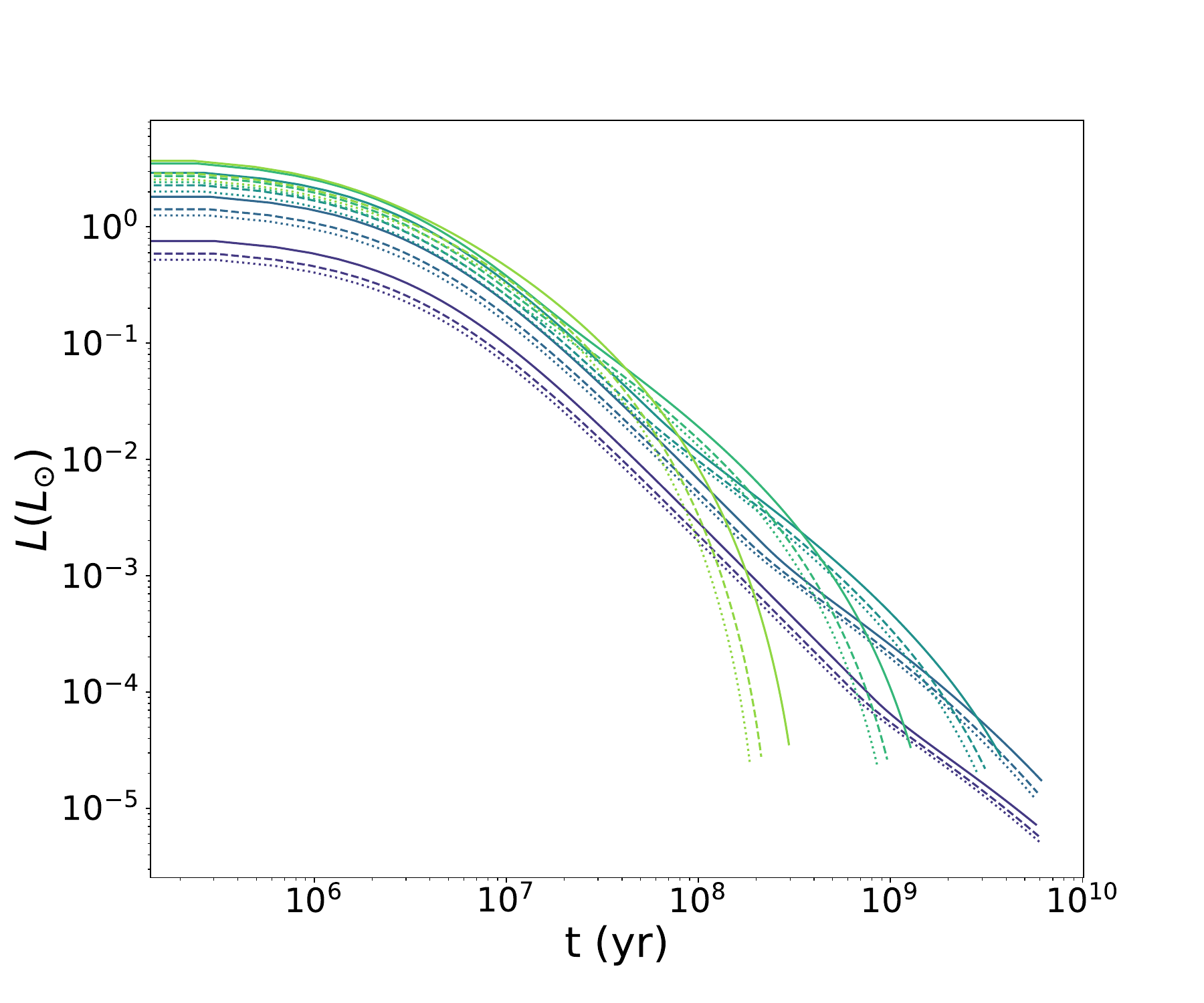}
\caption{Luminosity $L$ as a function of time $t$ for chameleon-screened WDs. We consider the chameleon model \eqref{eq:potential_&_conformal_coupling} with $n=1$, $\Lambda=10^{-18}\,M_P$, and $\beta = 0.1$ (dotted) and $\beta = 0.05$ (dashed). We also display the results for Newtonian gravity WDs (solid). Colours from purple to green indicate increasing central densities, namely   $\Tilde{\rho}_0 = 4.5\times10^{5},\,5.5\times10^{6},\,6.7\times10^{7},\,8.2\times10^{8},\,1.0\times10^{10}\text{ g}\,\text{cm}^{-3}$. The higher the coupling strength $\beta$, the faster the WD cools down.} \label{fig:Lvst_L1e18}
\end{figure}

The dependence of the mean specific heat on temperature for carbon WDs (recall Sec.~\ref{subsec:EoS}) is displayed  in Fig.~\ref{fig:c_VvsT_L1e18}. We observe that the $\Bar{c}_V-T$ curves in our ST theory are shifted to higher temperatures as compared to those in Newtonian gravity. This means that, for any given temperature $T$ below the temperature of the $\Bar{c}_V$ maximum, the specific heat is lower for chameleon-screened WDs. This also happens when the specific heat is dominated by the constant contribution of $c_V^{\text{ion}}=3/2$ (notice the drop in all curves in Fig.~\ref{fig:c_VvsT_L1e18}). However, between the $\Bar{c}_V$ maximum and this drop, the specific heat for WDs in our chameleon model is higher than for their purely Newtonian counterparts. Nevertheless, the overall values of the specific heat are smaller in our ST theory, as can be appreciated by looking at the maximum of each curve: the higher the coupling strength $\beta$, the lower the $\Bar{c}_V$ maximum. This leads to a faster WD cool-down. To see this explicitly, let us consider the luminosity $L$ of WDs, 
\begin{equation}\label{eq:luminosity}
    L=-\frac{M}{Am_p}\Bar{c}_V\frac{dT}{dt}\,,
\end{equation}
coming primarily from the thermal energy decrease of ions and electrons with respect to time $t$, with $M$ the stellar mass and $T$ the temperature of the star. To solve this equation, we use a luminosity fit  (see \cite{Koester_1990} and references therein for further examples)
\begin{equation}\label{eq:L/M}
    \frac{L}{M}=9.743\times10^{-21}T^{2.56}\frac{L_\odot}{M_\odot}\,,
\end{equation}
with the temperature $T$ in $K$. The associated dimming of WDs over time is displayed in Fig.~\ref{fig:Lvst_L1e18}, where we have assumed an initial temperature $T_\text{ini}=10^8\,K$ and allowed for cooling to a final temperature $T_\text{ini}=10^6\,K$. The period elapsed between these two points is the cooling time of the WD. We observe that the presence of the chameleon field makes the WDs cool faster, an effect that becomes more evident for higher densities. This indicates that the faster cooling is not entirely due to chameleon-screened WDs having lower masses. The different cooling rates displayed by low and high densities comes from the fact that the specific heat for less dense WDs is approximately constant for a significant range of temperatures, as seen in Fig.~\ref{fig:c_VvsT_L1e18}. Conversely, the strong temperature dependence that $\Bar{c}_V$ has in dense stars gives us the faster decay in Fig.~\ref{fig:Lvst_L1e18}.

\subsection{\label{subsec:ScalarProfi}Scalar Profiles}

\begin{figure*}[htbp]
    \centering
     \subfigure[]{%
        \includegraphics[width=0.45\linewidth]{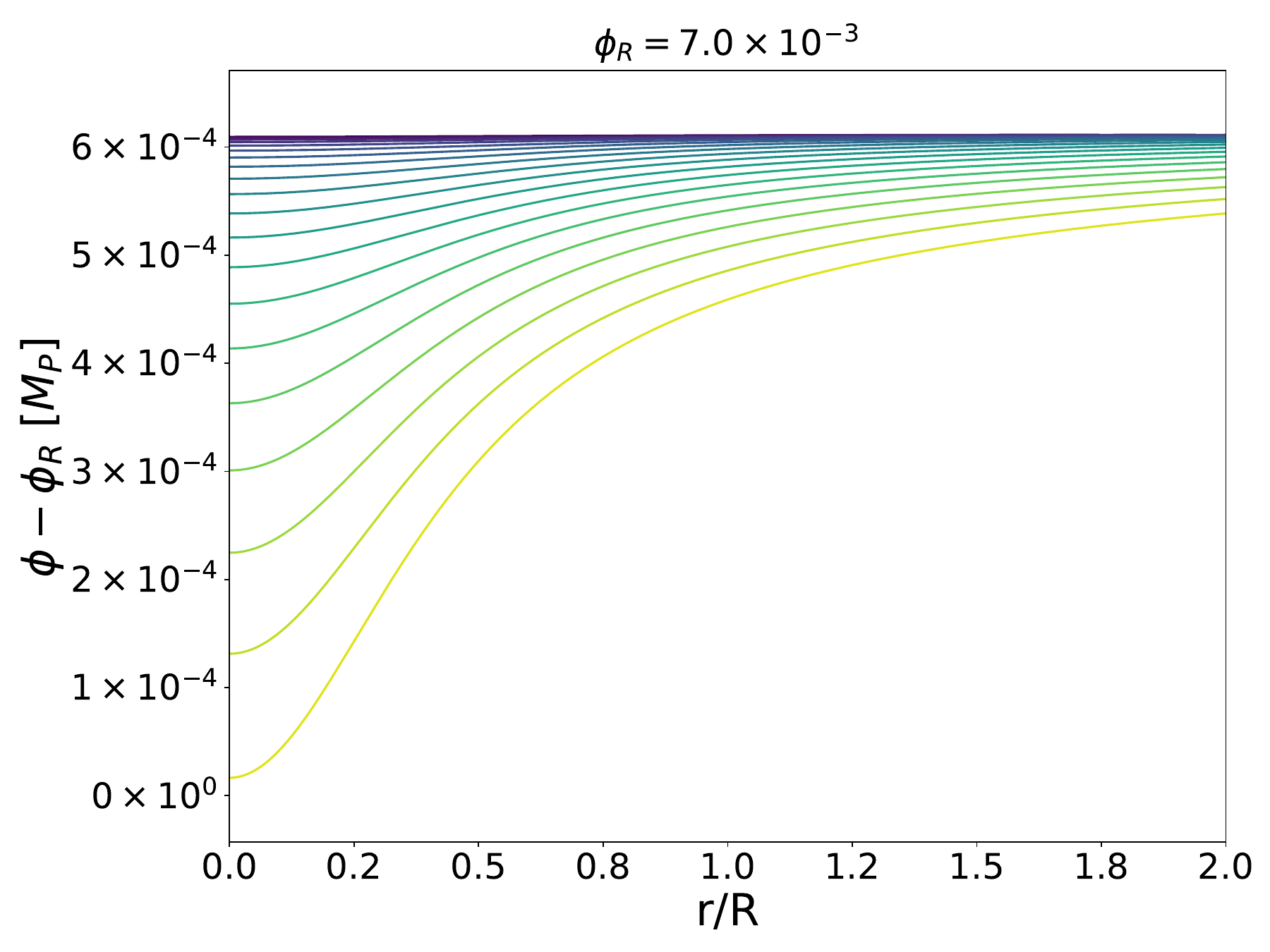}}
     \subfigure[]{
        \includegraphics[width=0.45\linewidth]{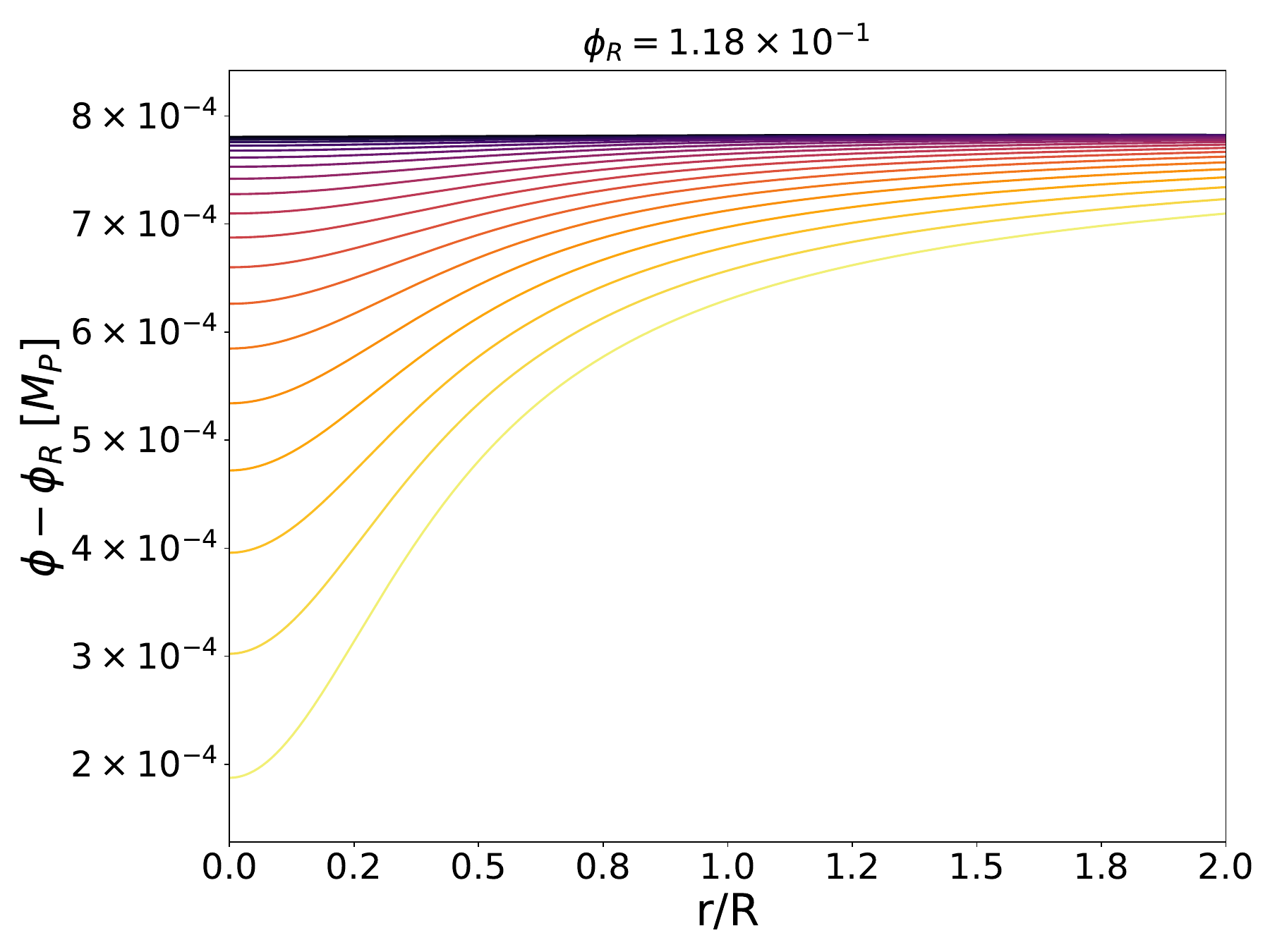}}
     \subfigure[]{%
        \includegraphics[width=0.45\linewidth]{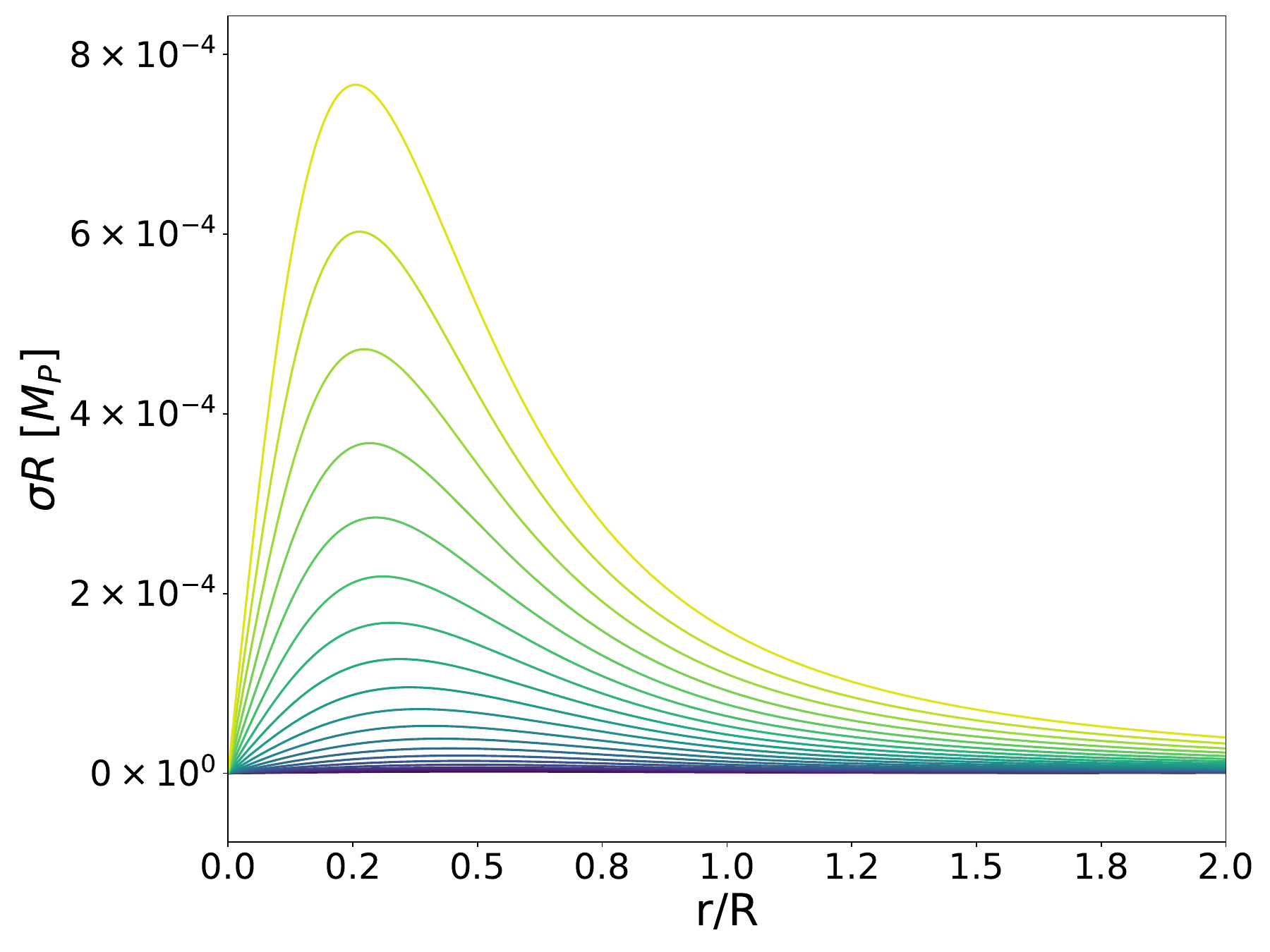}}
     \subfigure[]{
        \includegraphics[width=0.45\linewidth]{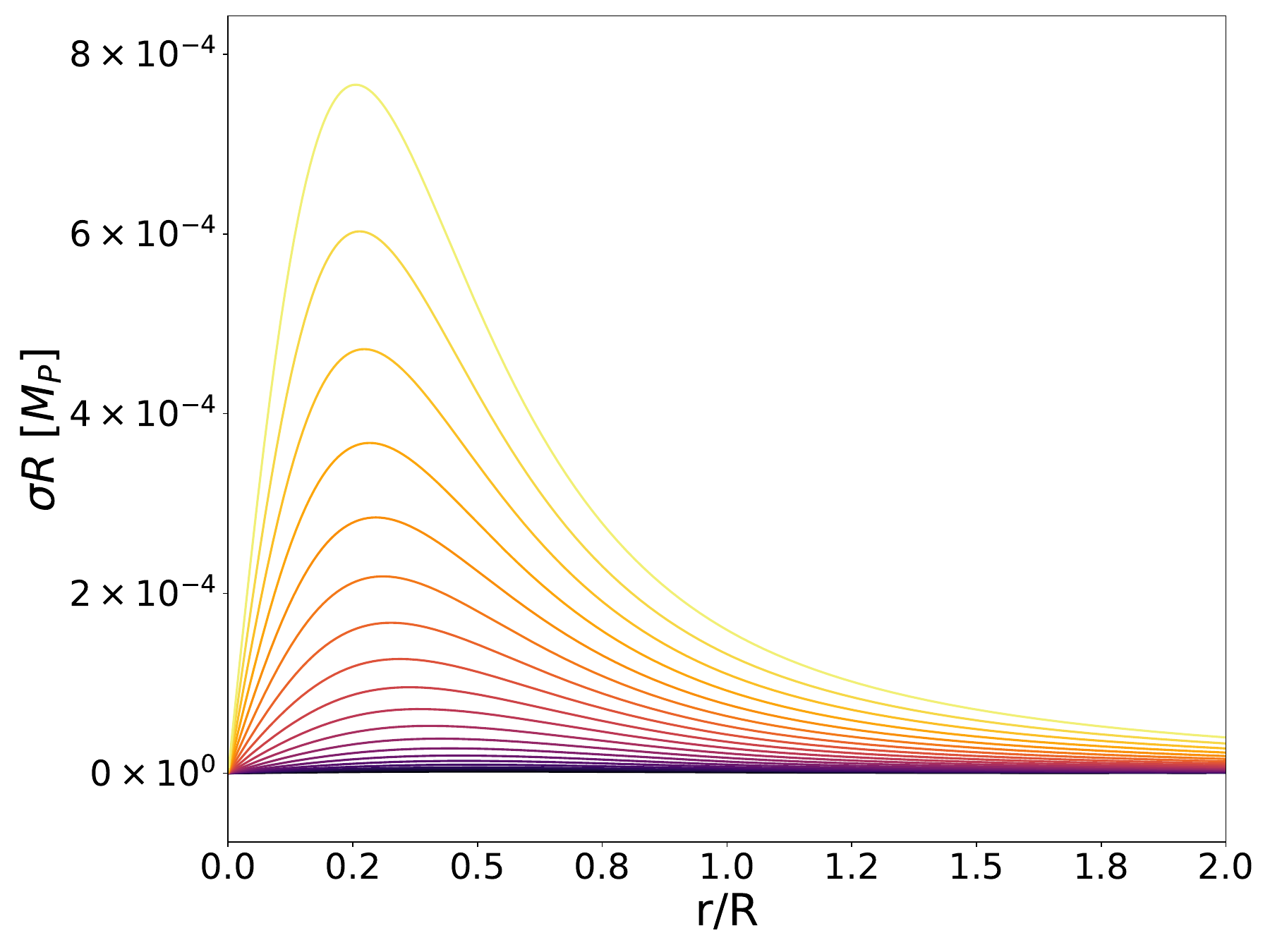}}
    \caption{\textit{Top panels}: Scalar field radial profiles. Notice that the labels indicate the difference between the scalar field and a reference value since the variation is small compared to the latter. \textit{Bottom panels}: Scalar field gradient radial profiles. \\
    The radial coordinate has been normalised to the respective stellar radius $R$ for each curve. 
    We consider the chameleon model \eqref{eq:potential_&_conformal_coupling} with $\Lambda=10^{-19}\,M_P$, $\beta = 0.05$, and $n=1$ (left panels, green tones) and $n=2$ (right panels, orange tones). Tones from dark to bright indicate increasing central densities from $\Tilde{\rho}_{0,\text{min}} = 7\times10^{4} \text{ g}\,\text{cm}^{-3}$ to $\Tilde{\rho}_{0,\text{max}} = 10^{10} \text{ g}\,\text{cm}^{-3}$.}\label{fig:phi&sigmavsrR_L5e19_a5e2}
\end{figure*}

\begin{figure*}[htbp]
    \centering
     \subfigure[]{%
        \includegraphics[width=0.45\linewidth]{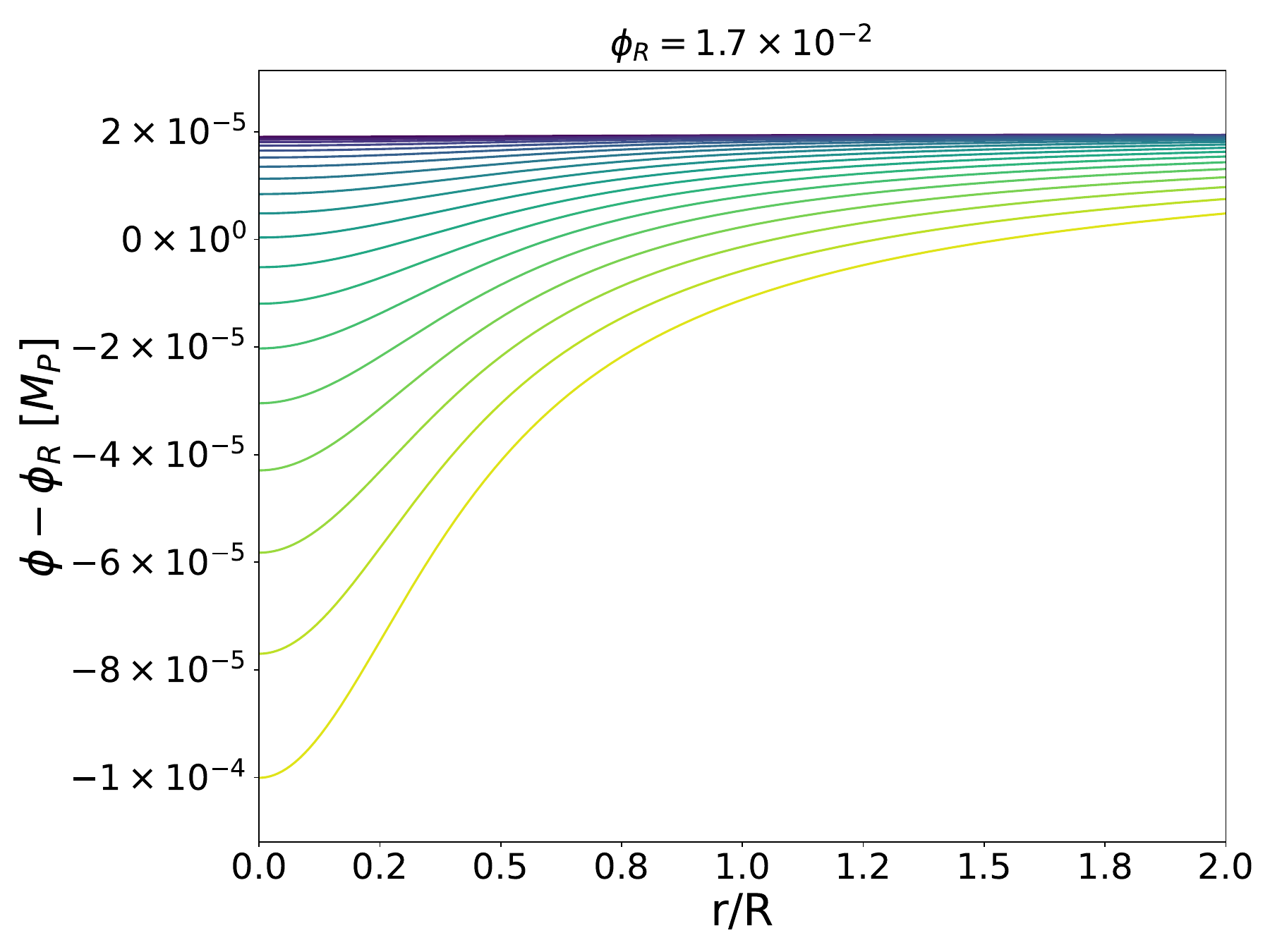}}
     \subfigure[]{
        \includegraphics[width=0.45\linewidth]{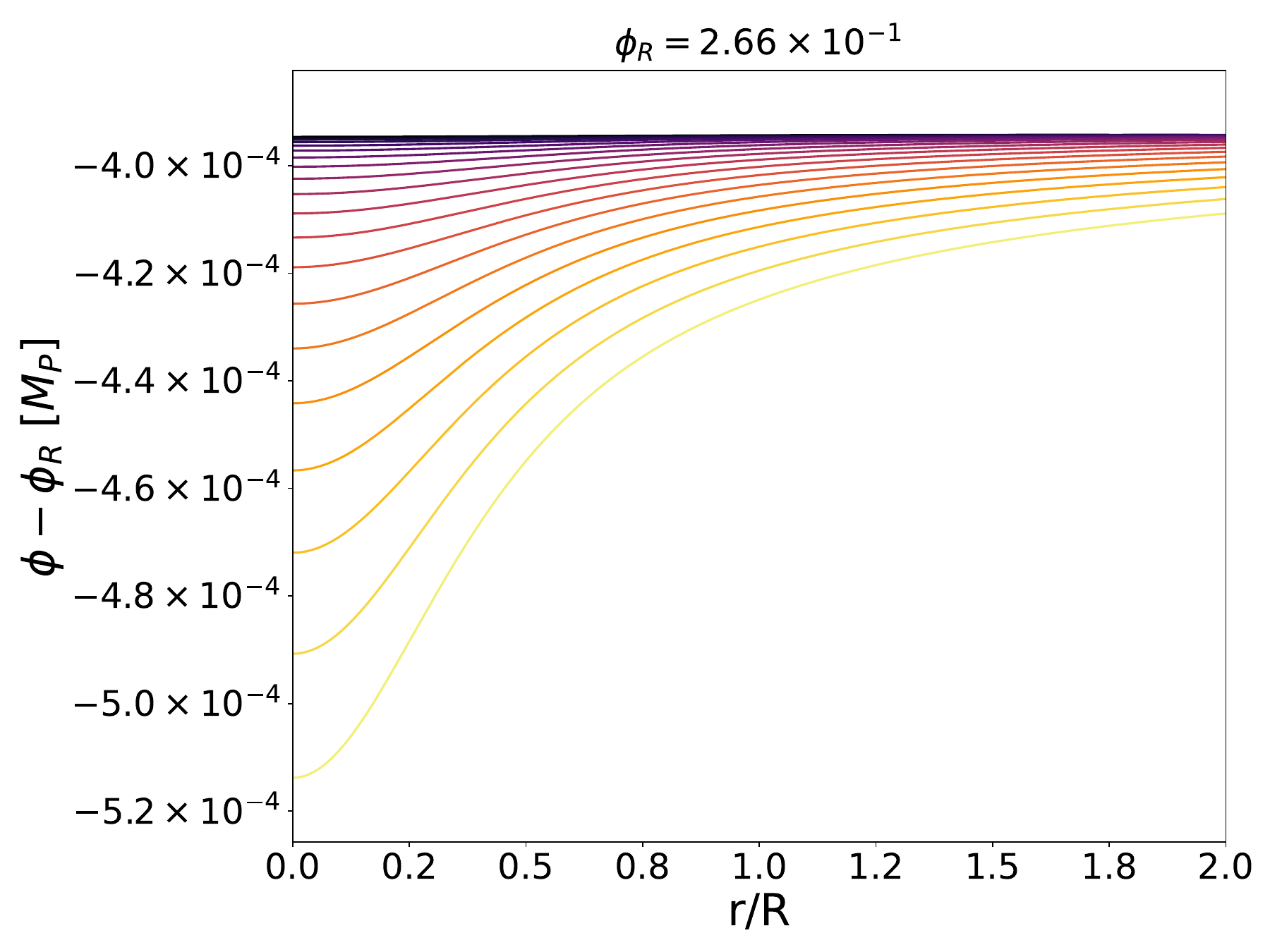}}
         \subfigure[]{%
        \includegraphics[width=0.45\linewidth]{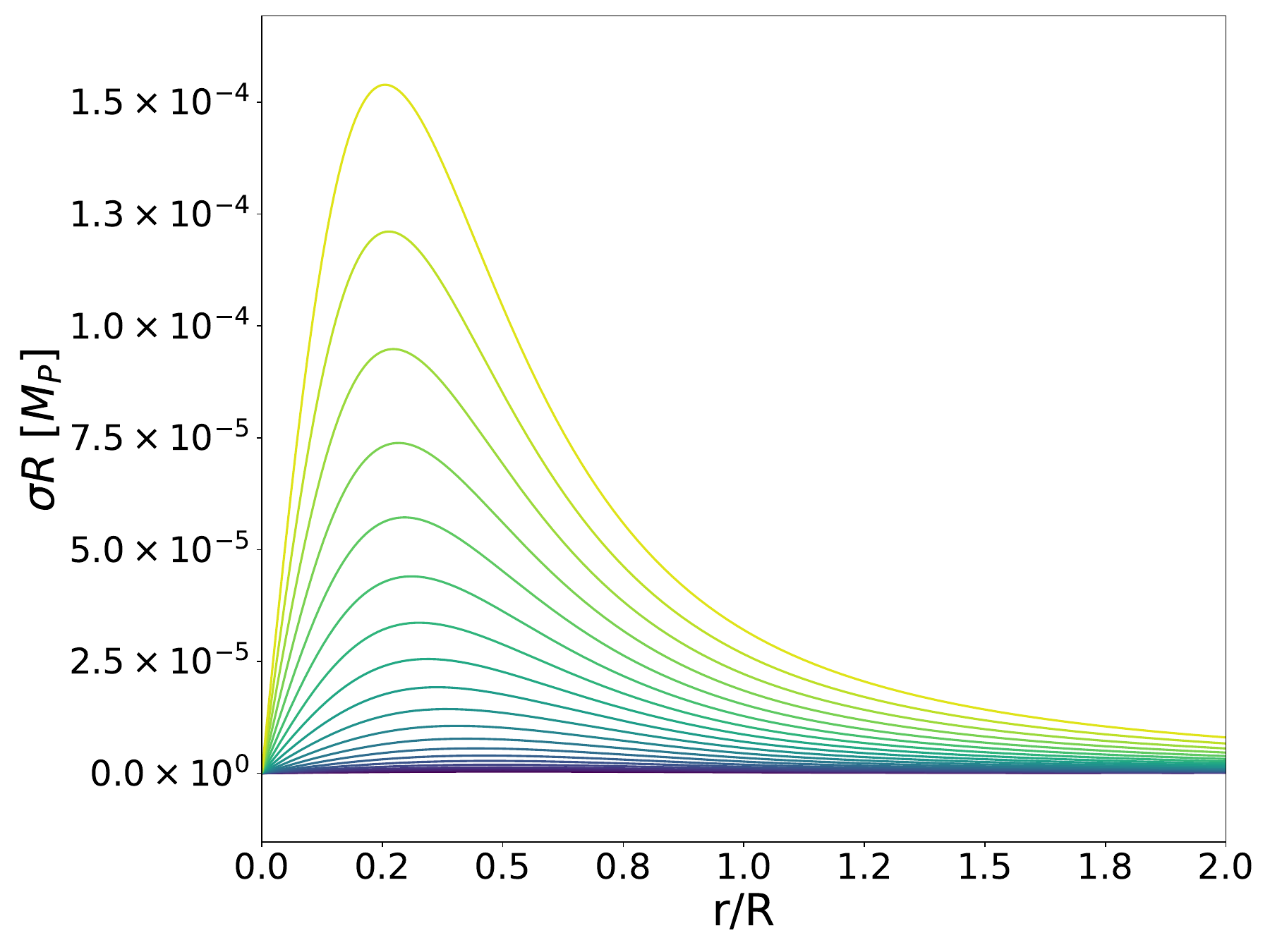}}
     \subfigure[]{
        \includegraphics[width=0.45\linewidth]{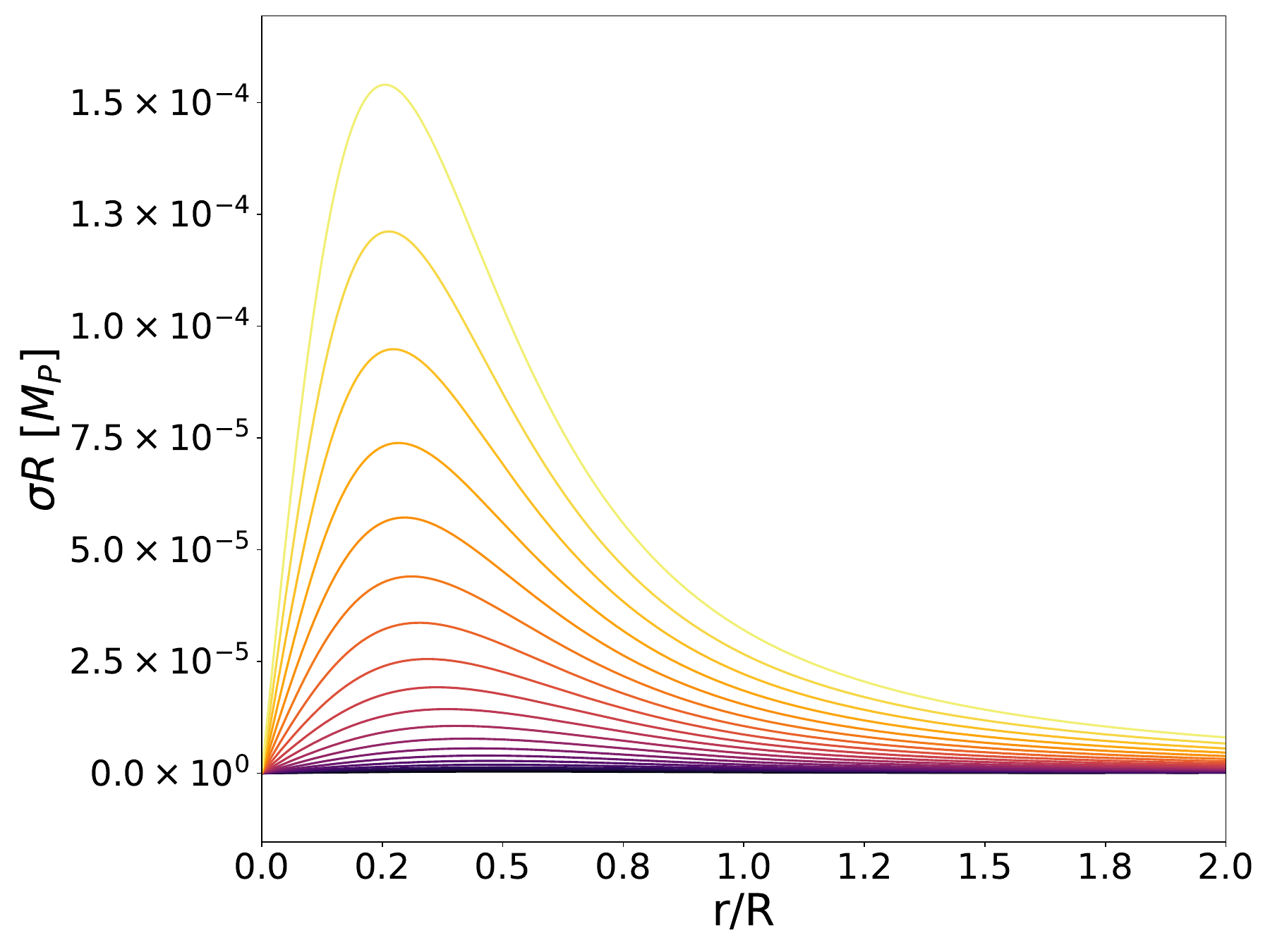}}
    \caption{\textit{Top panels}: Scalar field radial profiles. Notice that the labels indicate the difference between the scalar field and a reference value since the variation is small compared to the latter. \textit{Bottom panels}: Scalar field gradient radial profiles. \\
    The radial coordinate has been normalised to the respective stellar radius $R$ for each curve. 
    We consider the chameleon model \eqref{eq:potential_&_conformal_coupling} with $\Lambda=10^{-19}\,M_P$, $\beta = 0.01$, and $n=1$ (left panels, green tones) and $n=2$ (right panels, orange tones). Tones from dark to bright indicate increasing central densities from $\Tilde{\rho}_{0,\text{min}} = 7\times10^{4} \text{ g}\,\text{cm}^{-3}$ to $\Tilde{\rho}_{0,\text{max}} = 10^{10} \text{ g}\,\text{cm}^{-3}$.}\label{fig:phi&sigmavsrR_L5e19_a1e2}
\end{figure*}

In Fig.~\ref{fig:phi&sigmavsrR_L5e19_a5e2}, we display the radial profiles for the scalar field $\phi(r)$ and its gradient $\sigma(r)$ for two different parameter choices for the chameleon model \eqref{eq:potential_&_conformal_coupling}, namely $n=1$ and $n=2$ for fixed $\beta=0.05$ and $\Lambda=10^{-19}\,M_P$. In Fig.~\ref{fig:phi&sigmavsrR_L5e19_a1e2} we plot the same radial profiles but for  $\beta=0.01$. In all four cases, the scalar field profile is approximately flat for low central densities (darker colours). As the central density increases, the scalar field becomes suppressed inside the star, giving rise to the characteristic thin-shell pattern \cite{Khoury_2004a}. This is the essence of the chameleon screening mechanism, as we explained in Sec.~\ref{subsec:Screening}. 

It is worth mentioning the difference between scalar field values between $n=1$ and $n=2$ scenarios. In both cases -- see panels (a) and (b) in Figs.~\ref{fig:phi&sigmavsrR_L5e19_a5e2} and \ref{fig:phi&sigmavsrR_L5e19_a1e2} --, the scalar field in the $n=2$ model is approximately two orders of magnitude larger than in the $n=1$ model. Yet, this means that the chameleon potential $V(\phi)$ (recall \eqref{eq:potential_&_conformal_coupling}) has higher values in the $n=1$ model. For instance, for the scalar field values in Fig.~\ref{fig:phi&sigmavsrR_L5e19_a5e2}, we have that $V\sim10^{-92}\,M_P^4$ in the $n=1$ model and that  $V\sim10^{-113}\,M_P^4$ for $n=2$. For Fig.~\ref{fig:phi&sigmavsrR_L5e19_a1e2}, one has that $V\sim10^{-92}\,M_P^4$ for $n=1$ and $V\sim10^{-113}\,M_P^4$ for $n=2$. It seems that, for a fixed pair of $\beta$ and $\Lambda$, the scalar field potential is much more significant for $n=1$ than for $n=2$. This could lead us to think that the chameleon screening will affect much more the WD in the $n=1$ case. Still, one must not forget the gradient contribution, which, as discussed in Sec.~\ref{subsec:Stellar}, plays a crucial role in the stellar structure. 

The radially normalised scalar field gradient -- normalised in the sense that we have multiplied each scalar field gradient profile $\sigma(r)$ by the corresponding stellar radius $R$ -- displays similar values between the $n=1$ and the $n=2$ cases for the two considered coupling strengths (cf. the bottom panels of Figs.~\ref{fig:phi&sigmavsrR_L5e19_a5e2} and ~\ref{fig:phi&sigmavsrR_L5e19_a1e2}). We also encounter the same values in both realisations if we change the chameleon energy scale $\Lambda$. This coincidence is qualitatively explained by the approximate solution \eqref{eq:thick_inside} presented in Sec.~\ref{subsec:Screening}. In particular, from the top panels of Figs.~\ref{fig:phi&sigmavsrR_L5e19_a5e2} and \ref{fig:phi&sigmavsrR_L5e19_a1e2}, we realise that the scalar field changes throughout the whole stellar profile, being therefore in the thick-shell regime where the referred solution applies. This contrasts with solutions for more compact objects, such as neutron stars \cite{de_Aguiar_2020}, which exhibit a thin-shell behaviour \footnote{It should be emphasised that both solutions are based on numerically feasible parameters, hence the results are not necessarily transferable to real astrophysical objects.}.
Deriving Eq.~\eqref{eq:thick_inside}  with respect to $r$, 
\begin{equation}\label{eq:thick_sigma}
    \sigma(r)=\frac{\beta\rho_cr}{3M_P}\,,
\end{equation}
and taking into account that $\sigma R$ reaches its maximum somewhere between $R/4$ and $R/2$, we have 
\begin{equation}\label{eq:thick_sigma_xR}
    \sigma(xR) = \frac{\beta\rho_cxR}{3M_P}=\frac{\beta xM}{M_P4\pi R^2}=\frac{2\beta xM_P\Phi_c}{R}\,,
\end{equation}
with $x$ the corresponding fraction of $R$. Note that in deriving this expression, we have replaced the average density with the stellar mass and radius, $\rho_c\equiv 3M/(4\pi R^3)$, and introduced the Newtonian potential at the surface of the star, $\Phi_c\equiv M/(8\pi M_P^2R)$. Hence, in this approximation, the maximum of $\sigma R$ depends only on the coupling strength $\beta$ and the Newtonian potential $\Phi_c$, that is
\begin{equation}\label{eq:thick_sigma_final}
    \sigma(xR)R = 2xM_P\beta\Phi_c\,.
\end{equation}
Due to this result, which as explained in Sec.~\ref{subsec:Screening} is essentially based on neglecting the contribution of the potential as compared to the chameleon coupling function, we expected no dependence on the potential parameters $n$ and $\Lambda$. What we did not expect is that the numerical results would also be independent of such parameters. It must be said though that Eq.~\eqref{eq:thick_sigma_final} gives values an order of magnitude below the ones we have obtained computationally. Still, the ratio between both values of the $\sigma R$ maxima is consistent through all the realisations of the chameleon model that we have studied. We find this agreement between our results and the analytical solution noteworthy.

\subsection{\label{subsec:MassRadius}Mass-Radius Relation}

\begin{figure*}[]
    \centering
     \subfigure[]{%
        \includegraphics[width=0.48\linewidth]{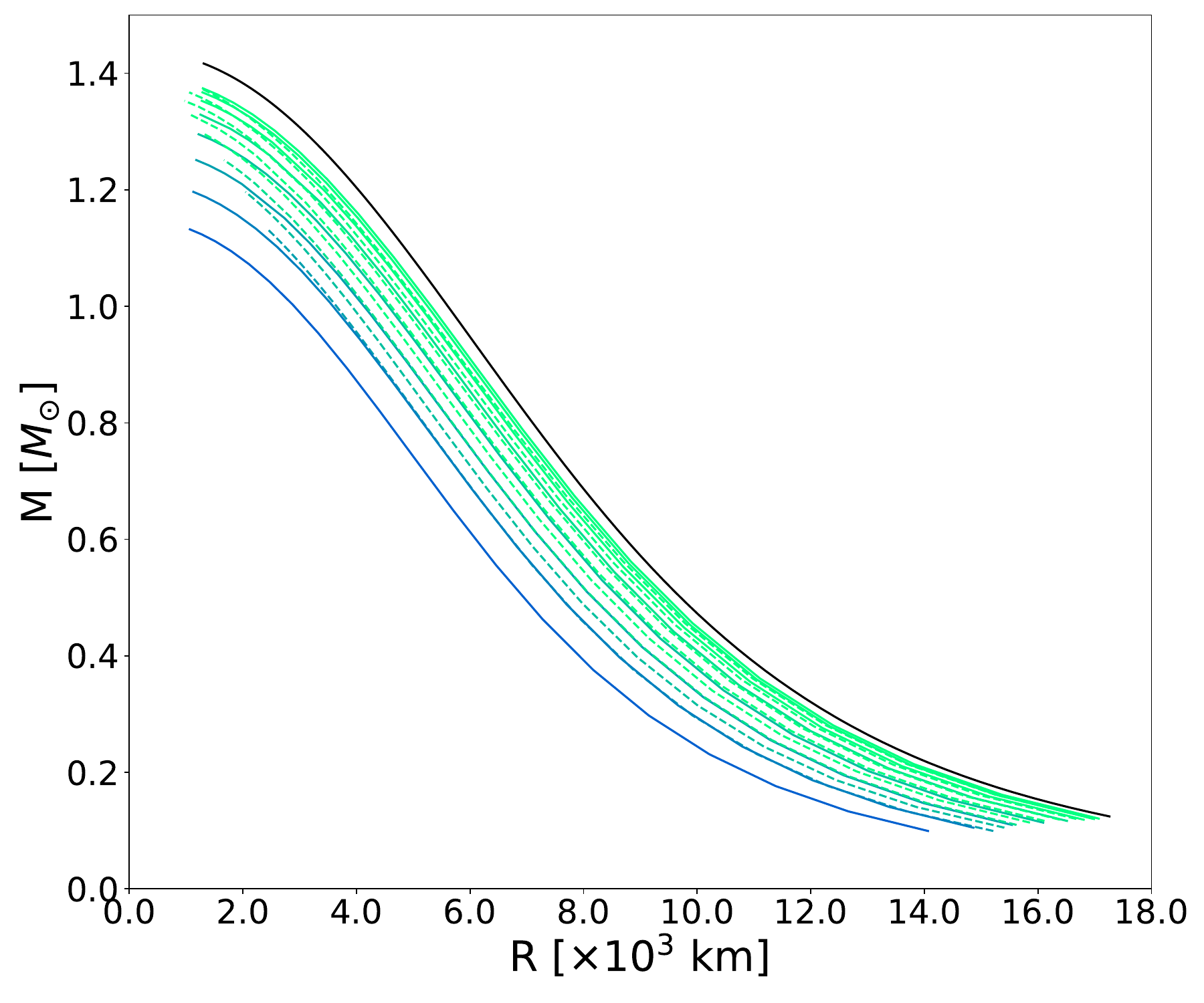}}
     \subfigure[]{
        \includegraphics[width=0.48\linewidth]{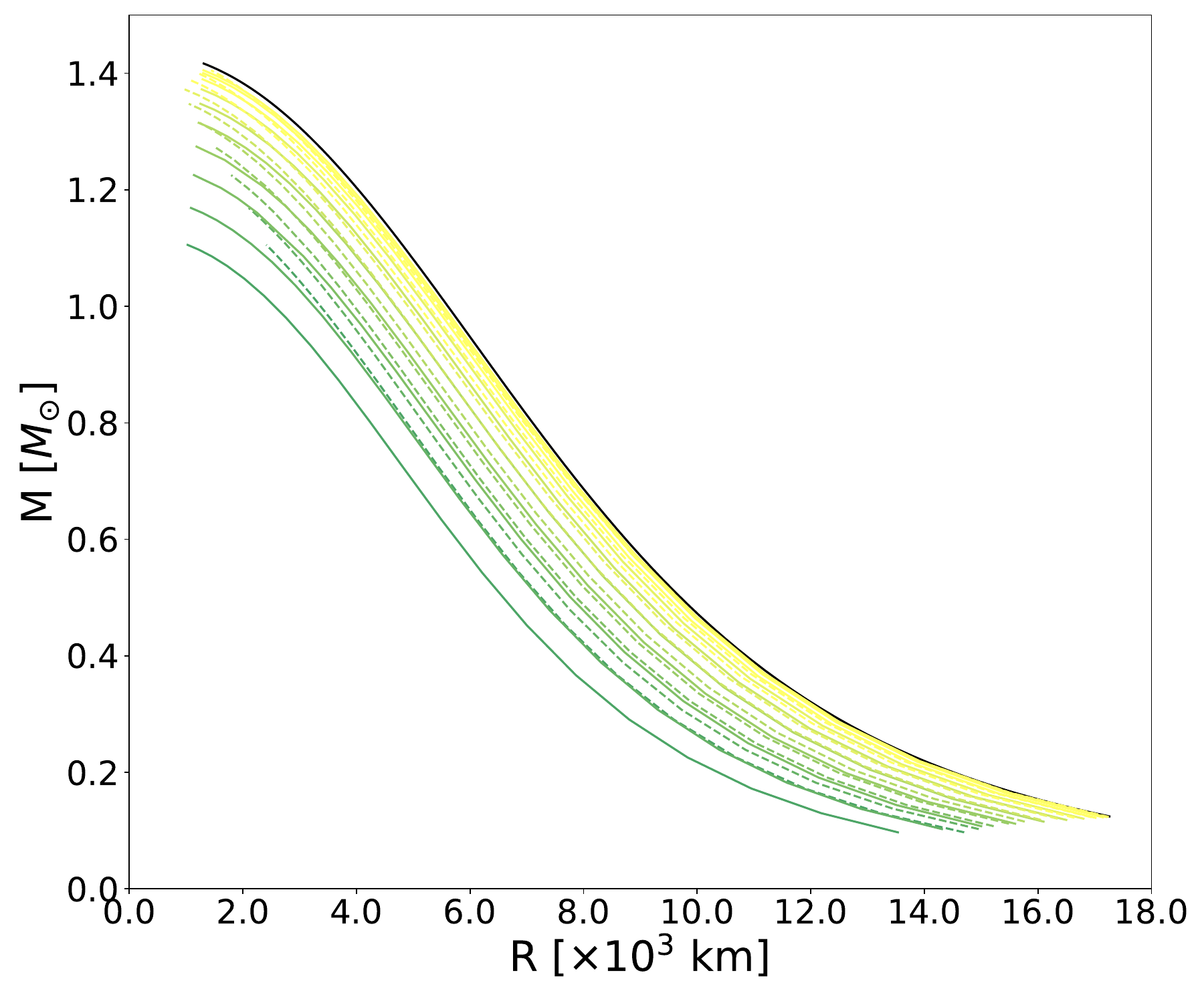}}
    \caption{Theoretical (solid) and parametric (dashed) MR curves for chameleon-screened WDs for $n=1$ and different values of $\beta$ and $\Lambda$, the latter changing in $10^{-19} M_P$ steps. We also include the WD MR curve in Newtonian gravity (black). \\ \textit{Left panel}: $\beta = 0.1$, $\Lambda = 1-8 \times 10^{-19} M_P$ (green to blue). \textit{Right panel}: $\beta = 0.05$, $\Lambda = 1-10 \times 10^{-19} M_P$ (yellow to green).}
    \label{fig:MR_curves}
\end{figure*}

Fig.~\ref{fig:MR_curves} contains numerically computed MR curves for chameleon realisations defined by the parameter values discussed at the beginning of this section, with $n=1$. As anticipated from the pressure and mass profiles in Fig.~\ref{fig:Pvsr_L1e18}, all MR curves of chameleon-screened WDs are below (or practically on top of) the MR curve predicted by Newtonian gravity. As expected, the smaller the coupling parameter $\beta$, the closer the MR curves for our ST theory are between them and to the Newtonian one. 

The MR curves for each explored value of $\beta$ tend to converge as $\Lambda$ decreases, suggesting they approach an asymptotic curve that never intersects the Newtonian one. This leads us to think that, if we were to consider much lower values of $\Lambda$, the MR curve for chameleon-screened WDs would never meet that for Newtonian WDs. Unfortunately, we need to significantly increase the numerical precision to explore such small energy scales.

Since the degeneracy between curves is evident -- in the sense that we could get the same masses and radii with various combinations of $n$, $\beta$, and $\Lambda$ -- it is useful to obtain a parametric function for them. A relation like $R=R(M,n,\beta,\Lambda)$ can bring order into chaos, allowing not only to compare different chameleon realisations between themselves but also between other models, such as other ST theories with screening mechanisms. Since all the solid curves in Fig.~\ref{fig:MR_curves} appear to have the same shape, we assume this formula for the mass-radius relation \cite{Greiner_1995}
\begin{equation}\label{eq:Fit_Parameters}
    R = R_*\left(\frac{M}{M_\odot}\right)^{-\frac{1}{3}}\sqrt{1 - \left(\frac{M}{M_*}\right)^\frac{4}{3}},
\end{equation}
which perfectly fits the MR curve for WDs in Newtonian gravity. In that case, we have that $R_*= 8.83\times10^3$ km and the $M_*$ parameter naturally coincides with the Chandrasekhar mass, i.e. $M_*=M_{Ch}=1.45$ $M_\odot$. 

To parameterise the MR curve of the chameleon-screened WDs, we turn the constants $R_*$ and $M_*$ into functions of the parameters of our chameleon model. After having explored the dependence of maximum radii and maximum masses with $\beta$ and $\Lambda$, we found that parabolic functions are a good ansatz, namely
\begin{eqnarray}
    R_*&=R_0+A_1\beta+A_2\beta^2+B_1\frac{\Lambda}{\Lambda_0}+B_2\left(\frac{\Lambda}{\Lambda_0}\right)^2\,,\label{eq:Fit_Formula_R}\\
    M_*&=M_0+C_1\beta+C_2\beta^2+D_1\frac{\Lambda}{\Lambda_0}+D_2\left(\frac{\Lambda}{\Lambda_0}\right)^2\,,\label{eq:Fit_Formula_M}
\end{eqnarray}
with $\Lambda_0=10^{-19}\,M_P$ a normalisation factor and $R_0$, $M_0$, $A_1$, $A_2$, $B_1$, $B_2$, $C_1$, $C_2$, $D_1$, $D_2$ constants. The best-fit values for these parameters are summarized in Tab.~\ref{tab:best-fit}, with a coefficient of determination $R^2= 0.992$. The corresponding curves are shown in Fig.~\ref{fig:MR_curves}, with the same colouring as their numerical counterparts. As apparent in these plots, the agreement between numerics and fitting formulae is sufficient for all practical purposes.

\begin{table}[h!]
    \centering
    \begin{tabular}{|c|c||c|c|}
        \hline
        Parameter & Value (km) & Parameter & Value ($M_\odot$) \\
        \hline
        $R_0$ & $9.638\times10^3$ & $M_0$ & $1.532$ \\
        \hline
        $A_1$ & $-2.005\times10^4$ & $C_1$ & $-1.919$ \\
        \hline
        $A_2$ & $1.066\times10^5$ & $C_2$ & $7.528$ \\
        \hline
        $B_1$ & $-1.434\times10^2$ & $D_1$ & $-1.179\times10^{-2}$ \\
        \hline
        $B_2$ & $-5.941$ & $D_2$ & $-1.080\times10^{-3}$ \\
        \hline
    \end{tabular}
    \caption{Best-fit values for the parabolic parameterization of MR curves for chameleon-screened WDs with $n=1$, defined by Eqs.~\eqref{eq:Fit_Parameters}, \eqref{eq:Fit_Formula_R}, and \eqref{eq:Fit_Formula_M}, with a $R^2= 0.992$ coefficient of determination. Note that $\beta\ll1$ results in the larger values for the $A_i$ and $C_i$ coefficients.}
    \label{tab:best-fit}
\end{table}

\section{\label{sec:Conclusion}Conclusion}

In this work, we have studied the effect that chameleon screening can have on the structure of WDs. They are auspicious candidates to test alternative theories of gravity since there is extensive observational data available and we have a sensible grasp of the EoS describing the matter within them. We have shown that the Newtonian approximation accurately describes WDs in this particular kind of ST theory, as it happens in GR. We have considered a Chandrasekhar EoS and solved the equilibrium equations with a custom-designed shooting method.

We have seen that the presence of the chameleon field affects the WD's internal pressure, causing it to drop prematurely as compared to when there is no scalar field. This leads to smaller stellar masses and radii which, in turn, shifts the MR curves below the MR relation predicted by Newtonian gravity. However, those stars above the theoretical curve -- which are the majority -- cannot come from any chameleon realisation, being therefore necessary to invoke other mechanisms such as strong magnetic fields \cite{Das:2014ssa,Roy:2019nja} or inverse chameleon settings \cite{Wei:2021xek} to explain them.

The existence of the chameleon field alters the specific heat of WDs too, lowering their values and reducing their cooling times. We have shown this effect to be more pronounced for denser stars and stronger $\beta$ couplings, confirming the role of the scalar field in the cooling process, which turns out to be not just a direct consequence of the smaller stellar masses of the chameleon-screened WDs.

We have examined the radial profiles of the scalar field and its gradient, considering a wide range of numerically feasible values for the model parameters -- the energy scale $\Lambda$ and the coupling strength $\beta$ -- and showing, for the first time in the literature, $n=2$ realisations of the chameleon. This allowed us to identify a similarity relation of the chameleon theory for the radially normalised scalar field gradient, that is $\sigma R$. In the thick-shell regime of the chameleon screening, the $\sigma R$ maximum is determined by $\beta$, and we have encountered the same behaviour in our numerical results, independently of the $\Lambda$ and $n$ values considered.

After exploring the ranges of the chameleon energy scales and coupling strengths, we have inferred parametric expressions for the MR relations that depend on the mentioned parameters. This result allows us to check for degeneracies between other classes of screening mechanisms once the analogous formulae are derived without having to numerically solve the ODE systems again, a computationally time-consuming task.

A continuation of this work could be to explore the stability of chameleon-screened WDs through radial perturbations, even studying the stable regime of WD oscillations. Furthermore, the effect that such a scalar field has on other properties of WDs, such as their crystallisation process, remains unexplored. Naturally, an interesting prospect would be to apply the presented framework and the developed computational tools to other ST theories.

\section*{Acknowledgments}
We are grateful to Raissa F. P. Mendes for detailed discussions on the numerical methods employed in this work. The numerical part of this work has been performed with the support of the Infraestrutura Nacional de Computação Distribuída (INCD), funded by the Fundação para a Ciência e a Tecnologia (FCT) and FEDER under the project 01/SAICT/2016 nº 022153. 
JBE (ORCID 0000-0003-4121-3179) and IL (ORCID 0000-0002-5011-9195) acknowledge the FCT, Portugal, for the financial support to the Center for Astrophysics and Gravitation - CENTRA, Instituto Superior Técnico, Universidade de Lisboa, through Project No. UIDB/00099/2020. JBE is grateful for the support of this agency through grant No. SFRH/BD/150989/2021 in the framework of the IDPASC-Portugal Doctoral Program.
IL also acknowledges FCT for the financial support through grant No. PTDC/FIS-AST/28920/2017. JR (ORCID 0000-0001-7545-1533) is supported by a Ramón y Cajal contract of the Spanish Ministry of Science and Innovation with Ref.~RYC2020-028870-I. This work was supported by the project PID2022-139841NB-I00 of MICIU/AEI/10.13039/501100011033 and FEDER, UE.

\appendix

\section{\label{app:TOVvsNWT}Relativistic and Newtonian Descriptions}

\begin{figure}[hbt!]
\includegraphics[width=\columnwidth]{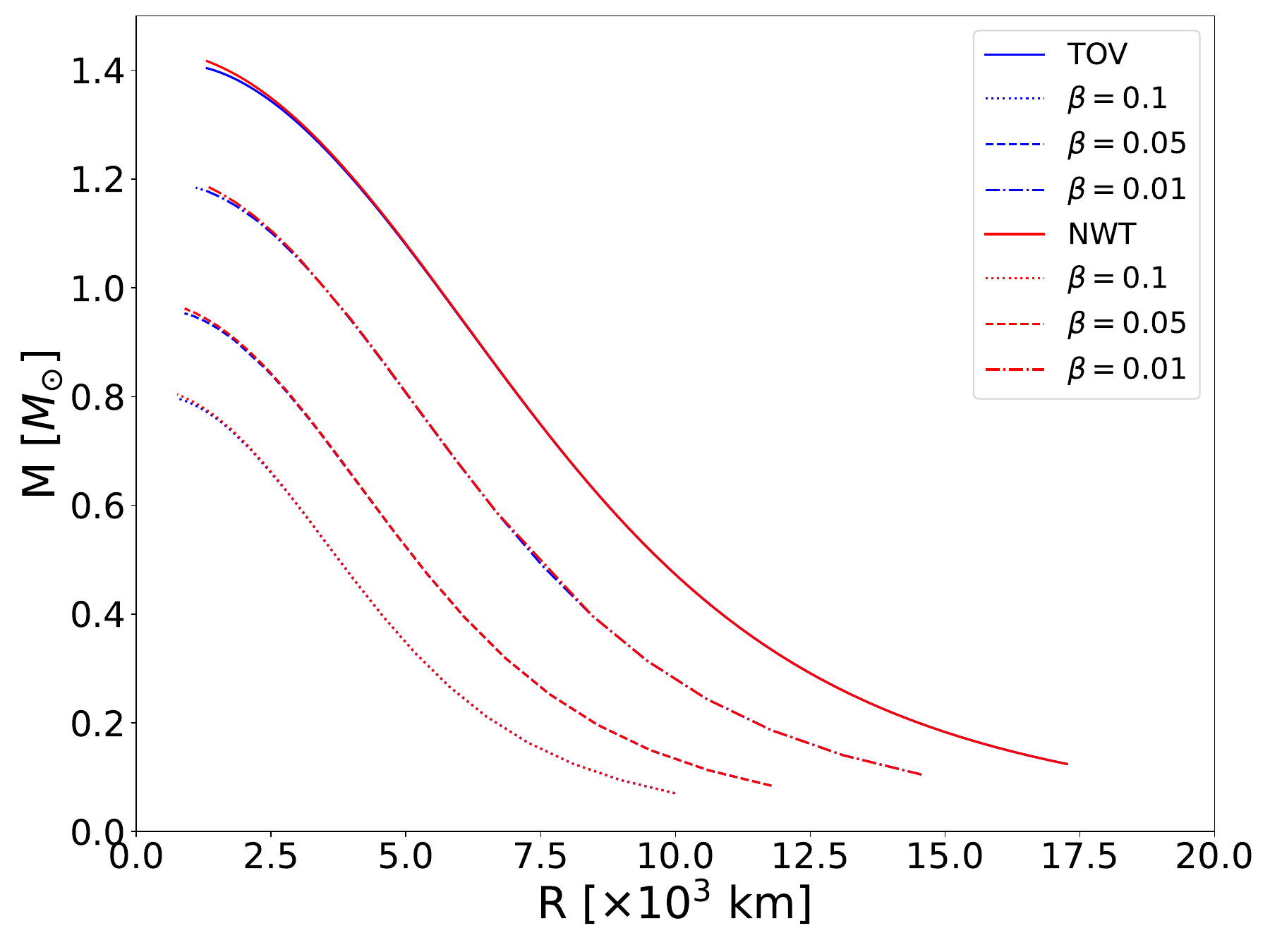}
\caption{\label{fig:MR_TOV&NWT_L12e18_a1e1_5e2_1e2}Theoretical MR curves for WDs in GR (solid) and in the presence of chameleon field for $n=1$, $\Lambda = 1.2 \times 10^{-18}\,M_P$ and $\beta = 0.1$ (dotted), $\beta = 0.05$ (dashed), and $\beta = 0.01$ (dash-dotted), both in the relativistic (TOV) approach (blue) and the Newtonian (NWT) approximation (red).}
\end{figure}

In this appendix, we compare the accuracy of the Newtonian description against the purely relativistic one for a static, spherically symmetric WD described by a perfect fluid energy-momentum tensor. We have already discussed the Newtonian approximation in Sec.~\ref{sec:Model}. For the relativistic description, we adopt the following line element to describe the spacetime
\begin{equation}\label{eq:metric_spherical_polar}
    ds^2=-e^{2\nu(r)}dt^2+e^{2\lambda(r)}dr^2+r^2d\Omega^2\,, 
\end{equation}
where $d\Omega^2=d\theta^2+\text{sin}^2\theta d\varphi^2$. Far away from the star, the spacetime must become Schwarzschild, that is
\begin{equation}\label{eq:metric_Schwarzschild-deSitter}
    ds^2=-f(r)dt^2+f(r)^{-1}dr^2+r^2d\Omega^2,
\end{equation}
where $f(r)=1-2a/r$. For the present case, we require that $a=m$ \cite{de_Aguiar_2020}, which means that 
\begin{equation}\label{eq:massTOV_potentials}
    \lambda(r)\rightarrow-\frac{1}{2}\text{ln}\left(1-\frac{2m(r)}{r}\right)
\end{equation}
for $r\gg R$, with $R$ being the stellar radius. Then, if we insert the metric of Eq.~\eqref{eq:metric_spherical_polar} into Eqs.~\eqref{eq:EF_eq}, \eqref{eq:EF_Matter_EOM}, and \eqref{eq:phi_eq}, 
we obtain the Tolman–Oppenheimer–Volkoff (TOV) equation with a scalar field contribution \footnote{Taking Eq.~\eqref{eq:background_nu} until the second term, replacing it in Eq.~\eqref{eq:background_p}, and setting $A(\phi)=1$, one recovers the well-known TOV equation.}
\begin{eqnarray}
    \nu'=&&\frac{1}{r-2m}\bigg\{\frac{m}{r}\nonumber\\
    &&+\frac{\kappa^2}{2}r^2\left[A^4\Tilde{P}+\left(\frac{1}{2}-\frac{m}{r}\right)\sigma^2-V\right]\bigg\},\label{eq:background_nu}\\
    m'=&&\frac{\kappa^2}{2}r^2\left[A^4\Tilde{\epsilon}+\left(\frac{1}{2}-\frac{m}{r}\right)\sigma^2+V\right],\label{eq:background_m}\\
    \Tilde{P}'=&&-(\Tilde{P}+\Tilde{\epsilon})\left(\nu'+\frac{A_{,\phi}}{A}\sigma \right),\label{eq:background_p}\\
    \phi'=&&\sigma,\label{eq:background_phi}\\
    \sigma'=&&\left[\frac{1}{r-2m}\left(m'+3\frac{m}{r}-2\right)-\nu'\right]\sigma\nonumber\\
    &&+\frac{r}{r-2m}\left[V_{,\phi}-A_{,\phi}A^3(3\Tilde{P}-\Tilde{\epsilon})\right],\label{eq:background_sigma}
\end{eqnarray}
where primes denote derivatives with respect to $r$. Once one chooses the model functions $V(\phi)$ and $A(\phi)$, and a suitable EoS, this ODE system can be numerically integrated, as it happened for Eqs.~\eqref{eq:background_Phi_Newton}-\eqref{eq:background_sigma_Newton}. To obtain a MR curve, we follow the same procedure we explained in the main body of the work (see Sec.~\ref{subsec:BC}). 

In Fig.~\ref{fig:MR_TOV&NWT_L12e18_a1e1_5e2_1e2}, we display the MR curve for several realisations of the chameleon model as well as for GR. Specifically, we have integrated Eqs.~\eqref{eq:background_nu}-\eqref{eq:background_sigma} and Eqs.~\eqref{eq:background_Phi_Newton}-\eqref{eq:background_sigma_Newton} for $n=1$, $\Lambda = 1.2 \times 10^{-18}\,M_P$, and different values of $\beta$. We have considered $\Lambda=\beta=0$ too, which effectively reduces the two ODE systems to the TOV equation and its Newtonian counterpart, respectively. As discussed in Sec.~\ref{subsec:EoS}, WDs are non-relativistic objects, and it is well-known that a Newtonian approximation is more than sufficient to describe them. Consequently, we were expecting what we see from the two solid curves in Fig.~\ref{fig:MR_TOV&NWT_L12e18_a1e1_5e2_1e2}, which show us that the masses and radii predicted by the TOV equation (in blue) coincide with those calculated in the Newtonian limit (in red). 

We observe a small discrepancy just in the more massive end of the curve, close to the Chandrasekhar limit of $1.4\,M_{\odot}$, which is slightly surpassed in the Newtonian limit. One can say that Newtonian physics overestimates radii and underestimates surface gravity, thus demonstrating the significance of general relativistic effects in determining the physical properties of these compact stars, only for particularly massive WDs. For instance, in \cite{Carvalho_2018}, they found that the radius predicted by GR for a WD with a mass of 1.415 $M_\odot$ is approximately 33\% smaller than that calculated in Newtonian physics. Still, it should be noted that they consider a different EoS from ours.

Regarding the MR curves computed in our ST theory, we find the same behaviour: the discrepancy between GR and Newtonian gravity manifests itself in the most massive WDs. As a result, we have shown that the Newtonian description is completely sufficient for chameleon-screened WDs, thus it was no blunder from our side to only consider the latter in the main body of the paper. Fortunately, this boosted our work from a computational perspective since Eqs.~\eqref{eq:background_Phi_Newton}-\eqref{eq:background_sigma_Newton} are much simpler than Eqs.~\eqref{eq:background_nu}-\eqref{eq:background_sigma}, hence another reason in favour of having ignored the modified TOV equation.

\bibliographystyle{elsarticle-num} 
\bibliography{references}

\end{document}